\documentclass[prb, superscriptaddress, amsmath, amssymb, twocolumn, aps, showpacs]{revtex4-1} 
\usepackage{xcolor}
\usepackage{natbib}
\usepackage{mhchem}
\usepackage{longtable}
\usepackage{graphicx}
\usepackage[lofdepth,lotdepth,caption=false]{subfig}
\usepackage{overpic}
\usepackage{hyperref}
\usepackage[capitalize]{cleveref}
\usepackage{verbatim}

\newcommand{\1}{{\rm 1\hspace*{-0.4ex}%
\rule{0.1ex}{1.52ex}\hspace*{0.2ex}}}

\graphicspath{{/home/vijay/Research/thermoelectric/draft/figures/}{/home/vijay/Research/thermoelectric/draft/band-figures/}}

\begin{document}
\title{Modeling the band structure of the higher manganese silicides starting from \ce{Mn4Si7}}
\author{Vijay Shankar V.}
\affiliation{Department of Physics and Center for Quantum Materials , University of Toronto, 60 St.~George St., Toronto, Ontario, M5S 1A7, Canada}
\author{Yu-Chih Tseng}
\affiliation{Canmet Materials, Natural Resources Canada, Hamilton Ontario L8P 0A5 Canada}
\author{Hae-Young Kee}
\email{hykee@physics.utoronto.ca}
\affiliation{Department of Physics and Center for Quantum Materials , University of Toronto, 60 St.~George St., Toronto, Ontario, M5S 1A7, Canada}
\affiliation{Canadian Institute for Advanced Research, Toronto, Ontario, M5G 1Z8, Canada}
\begin{abstract}
 The higher manganese silicides (HMS), with the chemical formula MnSi$_x$($x \approx 1.73 - 1.75$), have been attracted a lot of attention due to their potential application as thermoelectric materials.  While the electronic band structures of HMS have been previously studied using first principle calculations, the relation between crystal structures of Mn and Si atoms and their band structures is not well understood. Here we study \ce{Mn4Si7} using first principle calculations and show that a half cell consisting of five Mn atoms is the essential building block for \ce{Mn4Si7}. Using this insight, we construct a minimal tight-binding model for \ce{Mn4Si7} and other HMS including \ce{Mn11Si19} and \ce{Mn15Si26}. The role played by the Si atoms and possible ways to achieve higher figure of merit are also discussed.  
\end{abstract}
\maketitle

\section{Introduction}

Understanding the relationship between crystal structure and material properties continues to be a fascinating endeavour. Complex crystal structures are often behind novel material phenomena, and intermetallic compounds with Nowotny Chimney Ladder (NCL) structures, often found in semiconducting silicides, are a intriguing example of this relation between structure and properties\citep{nowotny1970crystal}. 

Of the semiconducting silicides with NCL structure, an interesting class of compounds are the higher manganese silicides (HMS) which have potential applications in optoelectronic\citep{Bost1987} and thermoelectric applications\citep{kawasumi1981crystal}. While the HMS have been readily utilized for technological applications, there are several puzzles about their behaviour yet to be fully understood. The HMS exist in several different crystalline phases such as \ce{Mn4Si7}\citep{Gottlieb2003}, \ce{Mn11Si19}\citep{Schwomma1964}, \ce{Mn15Si26}\citep{Knott1967}, and \ce{Mn27Si47}\citep{Zwilling1973}.  All the compounds have a tetragonal crystal structure based on that of \ce{TiSi2}\citep{DeRidder1971} with almost identical $a$ axis lattice constants with unusually long $c$ axis lattice constants ranging from 17\AA{} in \ce{Mn4Si7} to 118\AA{} in \ce{Mn27Si47}. 

In terms of electronic properties, resistivity measurements characterize the HMS to be degenerate semiconductors with an exponentially decreasing resistivity below 500K suggesting a gap of around 0.4~eV \citep{nishida1972semiconducting,Krontiras1988}. Holes are the majority carries in Hall effect measurements\citep{kawasumi1981crystal,nishida1972semiconducting,Krontiras1988} and the band gaps reported have ranged from 0.4~eV to 0.9~eV. \citet{Migas2008} undertook a comprehensive study of the band structure and electronic properties of the HMS using density functional theory where they reported degenerate semiconducting behaviour for the HMS apart from \ce{Mn4Si7} which is insulating. Noticeably, in their calculations, the magnitude of the band gap is similar for different members and the band structure is quite similar as one proceeds from \ce{Mn4Si7} to \ce{Mn27Si47}. 

In this paper, we attempt to paint a unifying picture of the nature of the electronic band structure of the HMS. Our study is based on the understanding that the HMS are NCL phases, some of whose structural properties were studied earlier by Frederickson and co-workers\citep{Fredrickson2004}. We study the crystal structure of the HMS by analogy to the NCLs studied in \citet{Fredrickson2004} and show how the HMS \ce{Mn4Si7} is an underlying building block for the other HMS. We then perform band structure calculations on the stoichiometrically simplest member of the series viz \ce{Mn4Si7} to understand the nature of the electronic states near the Fermi level. The 14 electron rule\citep{Fredrickson2004a}, applied to the HMS enables us to understand the reason why the HMS apart from \ce{Mn4Si7} are degenerate p-type semiconductors. We then use both the structural aspects and the results of the band structure calculations to construct a minimal tight-binding model for \ce{Mn4Si7} which is relatively easily extended to model the band structure of the other HMS. This approach offers a systematic understanding of the different roles of Mn and Si orbitals. 

The paper is organized as follows: in \cref{sec:crystal-structure}, we describe aspects of the crystal structure of \ce{Mn4Si7} and extend some of the earlier work on the crystal and electronic structure of the NCLs to the case of HMS. In particular, we highlight how \ce{Mn4Si7} is the structural building block for \ce{Mn11Si19}, \ce{Mn15Si26}, and \ce{Mn27Si47}. Then in \cref{sec:ab-initio}, we undertake a detailed study of the electronic structure of \ce{Mn4Si7} using the full-potential linearly augmented plane-wave (FP-LAPW) Elk\citep{elk} code. Then in \cref{sec:Tight_binding_model_with_five_sublattices}, we describe our tight-binding model for the band structure of \ce{Mn4Si7} and show how structural arguments discussed in \cref{sec:crystal-structure} enable us to extend our model for the other members of the HMS. Finally, we conclude in \cref{sec:Conclusions}.

\section{Crystal structure of higher manganese silicides}
\label{sec:crystal-structure}
\begin{figure}[h!]
\centering
\begin{overpic}[width=0.8\columnwidth]{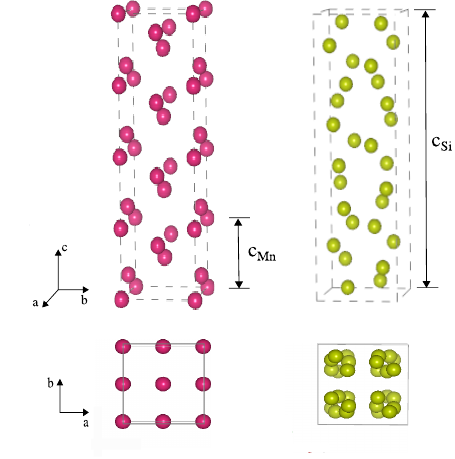}
\put(15,95){(a)}
\put(60,95){(b)}
\put(30,0){(c)}
\put(75,0){(d)}
\end{overpic}
\caption{(Color online) The chimney and ladder structures of \ce{Mn} and \ce{Si} respectively in \ce{Mn4Si7}. View along the $c$-axis in (a) and (b) show the four-fold and seven-fold helices of Mn (in pink) and Si (in light green) atoms respectively. The side views in (c) and (d) shows the chimney of \ce{Mn} atoms and the ladder of \ce{Si} atoms. The mismatch in the c-axis lattice periodicity of the \ce{Mn} and \ce{Si} atom gives rise to the long experimental $c$ axis lattice constant \label{fig:mn4si7-ncl}}
\end{figure}
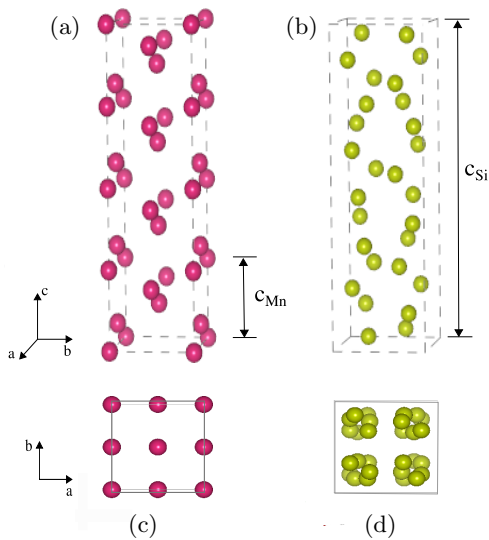

As mentioned in the introduction, HMS are compounds with the stoichiometric formula \ce{MnSi}$_x$ ($x \simeq 1.73 - 1.75$). The known phases all possess a tetragonal crystal structure derived from that of \ce{TiSi2}\citep{Ye1986} with similar $a$ axis lattice constants and long $c$ axis lattice constants. 

The HMS have a NCL structure\citep{nowotny1970crystal}, commonly found in intermetallic compounds formed between transition metal elements (T) and main group elements (E). In these chimney ladder phases, the T atoms form a tetragonal sublattice (the chimney) and the E atoms form a helix (the ladder). The T atoms form a square lattice when viewed along the the $c$-axis, which are actually four-fold helices with a pitch c$_{T}$ along the $c$-axis as shown for \ce{Mn} atoms in \ce{Mn4Si7} in \cref{fig:mn4si7-ncl}(a, c). The helix formed by the E atoms have different periodicities in different NCLs, and in the case of \ce{Mn4Si7}, they form seven-fold helices as shown in \cref{fig:mn4si7-ncl}(b, d). 

An additional structural feature of HMS first observed in electron diffraction patterns by \citet{DeRidder1971}, is the existence of regularly spaced satellite peaks\citep{Ye1986} which arise due to the mismatch between the lattice constants of \ce{Mn} and \ce{Si}.
These satellite peaks are most clear along the [110] direction with the spacing of the satellite peaks in reciprocal space, c$_{\text{pseudo}}^{*}$ related to the stoichiometry of the HMS phase Mn$_t$Si$_m$ as $(2t -m)c^{*} =c_{\text{pseudo}}^{*}$, where $c^{*}$ is the reciprocal lattice vector. 

Fredrickson et al.\citep{Frederickson2005Nowotny,Fredrickson2004} explained this relationship between the spacing of the satellite peaks and the main reciprocal lattice peaks by considering the real space structure of the NCL compunds. In real space, the relationship between the peaks translates to $(2t -m)c_{\text{pseudo}}=c$. By viewing these compounds along the [110] direction, as shown for the HMS in \cref{fig:HMS-110}, we see that $(2t-m)$ repeats of a unit with periodicity c$_{\text{pseudo}}$ make up the lattice vector $c$. In the case of the HMS, the pseudoperiodicity c$_\text{pseudo}$ is equivalent to the $c$ axis lattice constant of \ce{Mn4Si7}. In the next subsection, we provide some details of the structure of \ce{Mn4Si7} and elaborate how c$_{\text{pseudo}}$ arises in this compound. 

\subsection{\ce{Mn4Si7} structure} 
\label{sub:mn4si7-structure}

\begin{figure}[h!]
\centering
\begin{overpic}[width=\columnwidth]{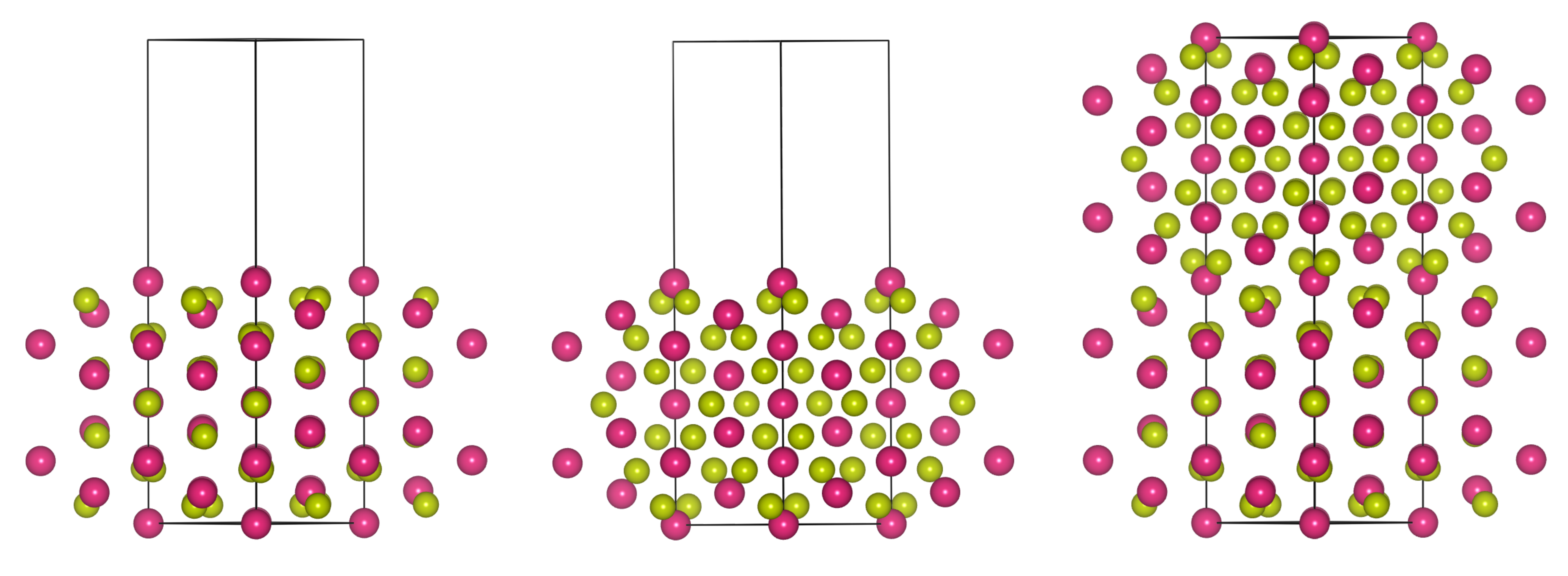}
\put(15,-2.2){(a)}
\put(50, -2.2){(b)}
\put(85,-2.2){(c)}
\end{overpic}
\caption{(Color online) Viewed along the [110] direction, \ce{Mn4Si7} is seen to be made up of two halves that are related to each other by a 90$^\circ$ rotation. The full unit cell is shown in (c), with the bottom half in (a) and a 90$^\circ$ rotated version of the bottom half in (b). The top half of \ce{Mn4Si7} is equivalent to the rotated version of the bottom half. As in \cref{fig:mn4si7-ncl}, Mn atoms are in pink and Si atoms are in light green. \label{fig:mn4si7-110}}
\end{figure}

\ce{Mn4Si7} comprises of Mn and Si helices with differing pitch as shown in \cref{fig:mn4si7-ncl}. Following \citet{Fredrickson2004}, we provide a brief description of the structure of \ce{Mn4Si7}. When viewed along the [110] direction, \ce{Mn4Si7} is seen to be made up of two half cells along the $c$ axis that are related to each other by a 90$^\circ$ rotation about the $c$-axis. The relatively long $c$-axis lattice constant in \ce{Mn4Si7} arises from the the juxtaposition of two slabs of a compound with stoichiometric formula \ce{MnSi2} (with a \ce{TiSi2} structure as in the case of \ce{RuGa2}\citep{Fredrickson2004}), that are rotated 90$^\circ$ w.r.t each other. While the \ce{Mn} atoms with four-fold rotational symmetry are unaffected by this rotation, the combination of these slabs will give rise to some \ce{Si} atoms that are unphysically close to each other near the interface. All the \ce{Si} atoms that are exactly at the interface are vacated and the remaining atoms relax their positions giving rise to the observed structure. 

In \ce{Mn4Si7}, we have two repeats of a half cell with eight \ce{Mn} and sixteen \ce{Si} atoms. The combination results in two interfaces -- one halfway along the $c$ axis and one at the end of the unit cell with two Si atoms lost at each interface. From the starting configuration of thirty two \ce{Si} atoms, four are removed giving rise to a total of sixteen \ce{Mn} atoms and twenty eight \ce{Si} atoms with the stoichiometric formula \ce{Mn4Si7}. 

\subsection{Building HMS via stacking of \ce{Mn4Si7} half cells}
\label{ssub:HMS-structure-stacking}

\begin{figure}[h]
\centering
  \begin{overpic}[scale=0.15]{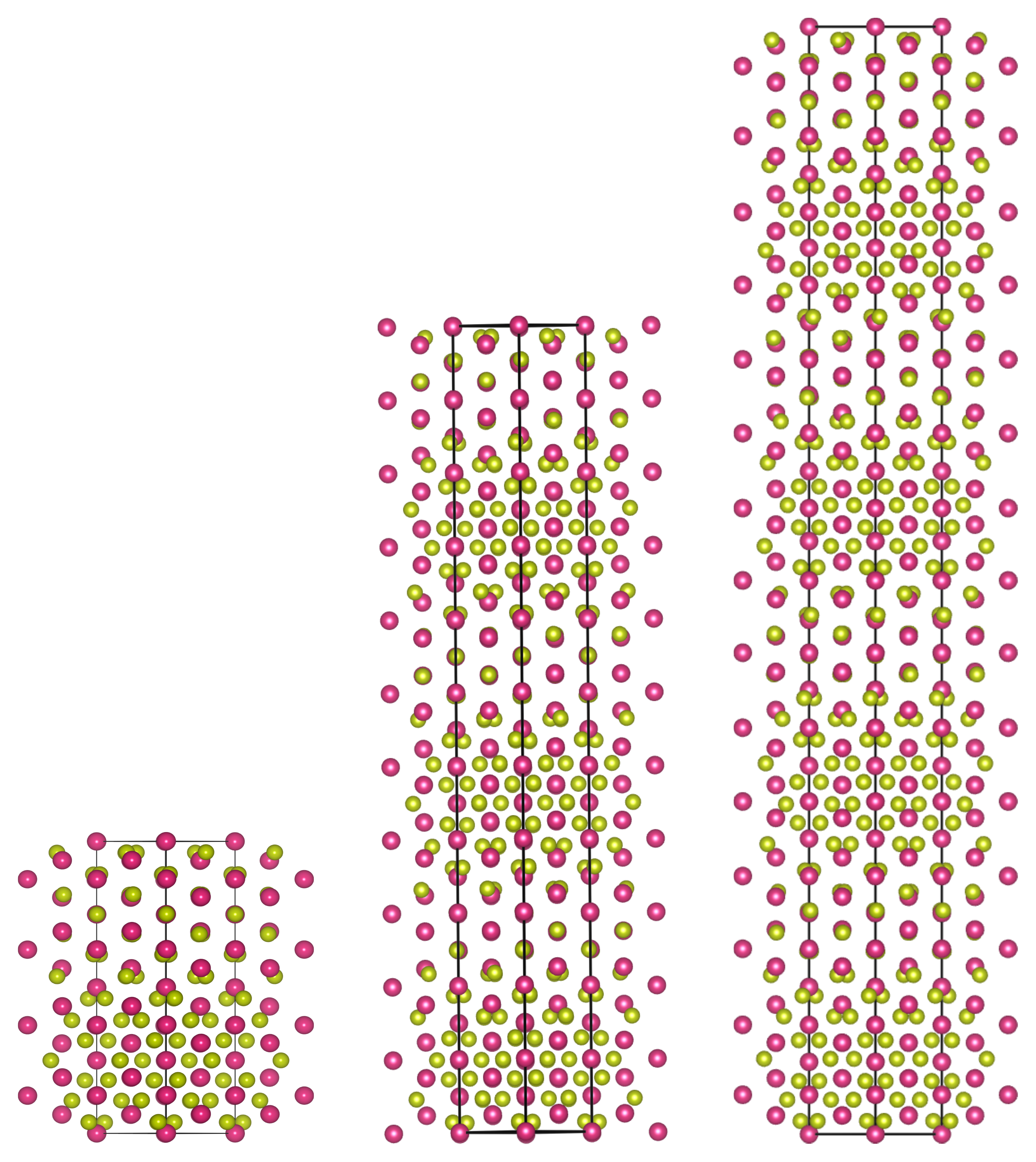}
  \put(5,-2){\tiny (a) \ce{Mn4Si7}}
  \put(35,-2){\tiny (b) \ce{Mn11Si19}}
   \put(64,-2){\tiny (c) \ce{Mn15Si26}}
  \end{overpic}
\caption{(Color online) HMS viewed along the [110] direction. In this projection, can see that \ce{Mn11Si19} and \ce{Mn15Si26} have an apparent periodicity that equals that of \ce{Mn4Si7}. The number of repeats of this c$_\text{pseudo}=c_{\text{\ce{Mn4Si7}}}$ satisfy the rule $(2t -m)c_{\text{pseudo}}=c$. For \ce{Mn11Si19}, $2t-m = (2*11 - 19) = 3$ and for \ce{Mn15Si26}, it is $2*15 - 26 = 4$, in agreement with the number of unit cells of \ce{Mn4Si7} that are stacked. 
\label{fig:HMS-110}}
\end{figure}

Within this picture, the other HMS with longer $c$-axis lattice constants can be built up by repeatedly stacking the half cell that makes up the unit cell of \ce{Mn4Si7} along the $c$-axis with a 90$^{\circ}$ rotation after every repeat. The unit cell of \ce{Mn11Si19} is made up of six half cells stacked along the $c$-axis, and that of \ce{Mn15Si26} and \ce{Mn27Si47} comprise of eight and fourteen half cells respectively. \cref{fig:HMS-110} shows the unit cells of \ce{Mn4Si7}, \ce{Mn11Si19}, and \ce{Mn15Si26} viewed from the [110] direction and the repeats of the \ce{Mn4Si7} units in the other two compounds are clearly visible.

In practice, the $c$-axis lattice constants of \ce{Mn11Si19}, \ce{Mn15Si26}, and \ce{Mn27Si47} are not integer multiples of the $c$-axis lattice constant of \ce{Mn4Si7}, even though the unit cells have an apparent periodicity equal to that of \ce{Mn4Si7}. The experimental $c$-axis lattice constants are smaller than the integer multiple obtained from the $(2t -m)$ rule \citep{Gottlieb2003,Schwomma1964,Knott1967,Zwilling1973}. That being said, there is still insight to be gained by viewing the longer HMS as stacked versions of \ce{Mn4Si7}, especially with regards to modeling the band structure as will be described in \cref{ssub:tb-stacking-hms}. With this background on the structural of HMS, we describe the results of our \emph{ab-initio} calculations for the band structure of \ce{Mn4Si7} in the next section.

\section{Electronic structure calculations}
\label{sec:ab-initio}
\begin{figure}[h!]
\centering
\subfloat[]{\label{fig:Mn4Si7-elk-band}\includegraphics[width=0.56\columnwidth]{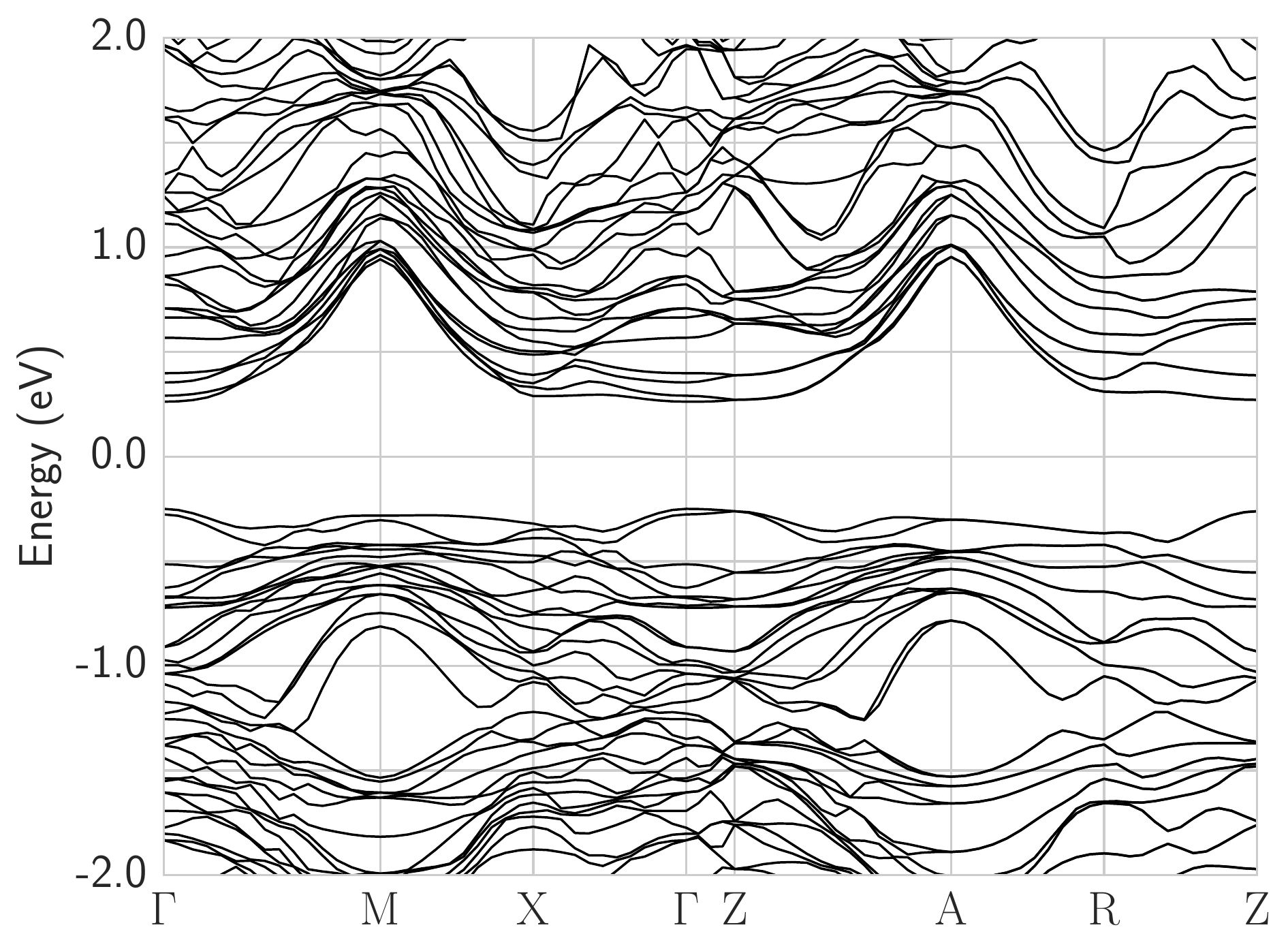}}
\hfill
\subfloat[]{\label{fig:Mn4Si7-elk-PDOS}\includegraphics[width=0.43\columnwidth]{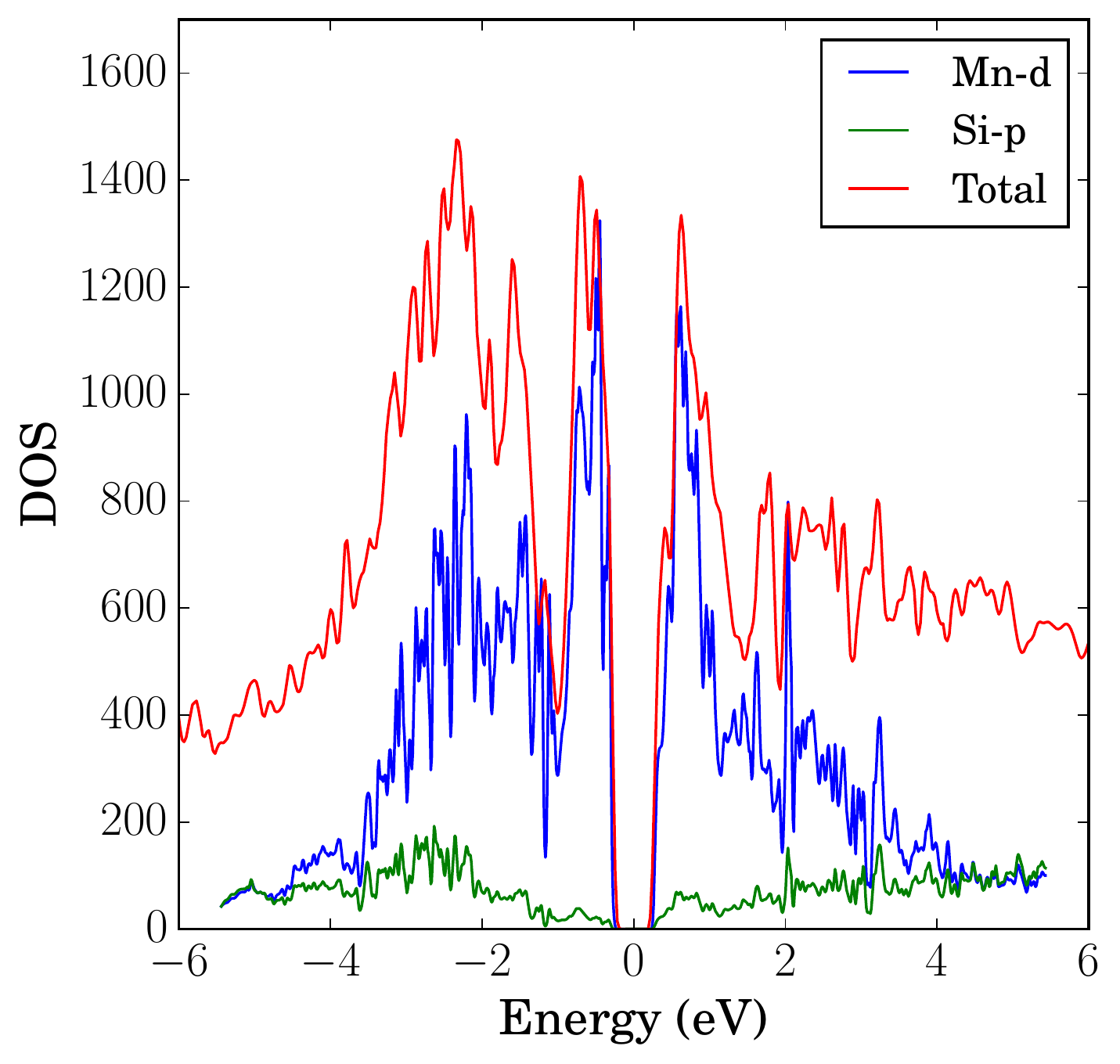}}
\caption{(Color online) Elk band structure (a) and DOS (b) for \ce{Mn4Si7}. The band structure (a) shows that the valence bands are hole like and are separated by an energy gap of 0.51 eV from the conduction band. The dispersion is quite similar in the $k_z$ = 0.0 ($\Gamma$-$X$-$M$-$\Gamma$) and $k_z$ = $\pi/2$ ($Z$-$A$-$R$-$Z$) planes. The total DOS (in red) has majority contribution from Mn $d$ orbitals (in blue). The partial density of states also shows that Si $p$ orbital contributions (in green) peak 2~eV below the Fermi level.
\label{fig:Mn4Si7-elk-band-DOS}}
\end{figure}

We computed the electronic band structure of \ce{Mn4Si7} using the Elk\citep{elk} package. Details of the computation are listed in \cref{app:computational}. The band structure shown in \cref{fig:Mn4Si7-elk-band} is insulating with a direct gap of 0.51~eV from the hole pocket below the Fermi level to the electron pocket at the $\Gamma$ point in the Brillouin zone  The dispersion in the $k_z =0$ plane ($\Gamma - X - M - \Gamma$) is much the same as that in the $k_z =\frac{\pi}{2}$ plane ($Z - A - R - Z$), due to the long $c$-axis lattice constant and results in a electronic structure that is practically two-dimensional in nature. The magnitude of the gap in our calculations also compares well with experimental measurements\citep{Gottlieb2003,nishida1972semiconducting}.

The density of states (DOS) in \cref{fig:Mn4Si7-elk-PDOS} is comprised of \ce{Mn} d-orbitals near the Fermi level and the \ce{Si} p - \ce{Mn} d bonding orbitals are clustered in a region 2-5~eV below the Fermi level. Similar density of states with the $d$-orbitals of the transition metal being the majority contributor to the states near the Fermi level has also been observed in band structure calculations for other NCL compounds\citep{Yannello2014}.

The band structure and the DOS agree qualitatively with previous calculations by \citet{Migas2008} with weakly dispersing valence bands below the Fermi level that come from the \ce{Mn} d orbitals. Quantitatively, they obtain a gap of 0.8~eV (see Fig.3 and Fig.5 in Ref.\citenum{Migas2008}) which is larger than our value of 0.5~eV. In experimental measurements, the band gap ranges from 0.42 to 0.98~eV\citep{Mahan2004potential,Gottlieb2003,nishida1972semiconducting,Teichert1996,Rebien2002}. \citet{Migas2008} also studied the band structures of \ce{Mn11Si19}, \ce{Mn15Si26}, and \ce{Mn27Si47} (see Fig.3 in Ref.\citenum{Migas2008}), where they obtained metallic ground states with the Fermi level lying within the valence bands. It was found that the dispersion near the Fermi level of these longer compounds is much like that of \ce{Mn4Si7} with the same energy gap from the top of the valence band to the bottom of the conduction band.  The band structures of these longer $c$-axis HMS show more pronounced two-dimensionality with almost identical dispersion in the $k_z = 0$ and $k_z = \frac{\pi}{2}$ planes. The similarity in the band dispersion, the magnitude of the energy gap, and the enhanced two-dimensionality of these HMS lend further credence to the idea that these HMS are best thought of as stacked version of \ce{Mn4Si7} as described in \cref{ssub:HMS-structure-stacking}. 

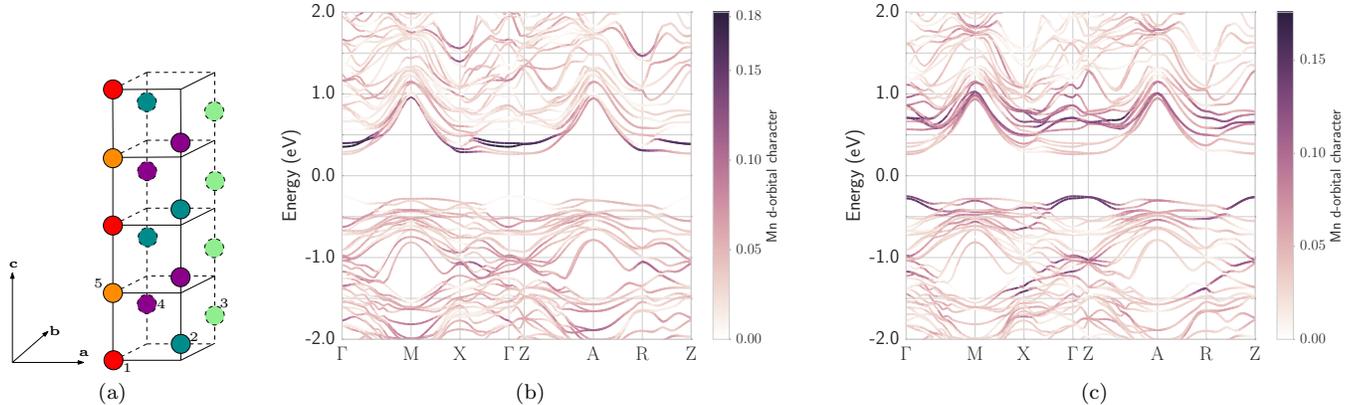
\begin{figure*}
\centering 
 \subfloat[]{\label{fig:unitcell-tight-binding-5-sites}
 \begin{overpic}[scale=0.4]{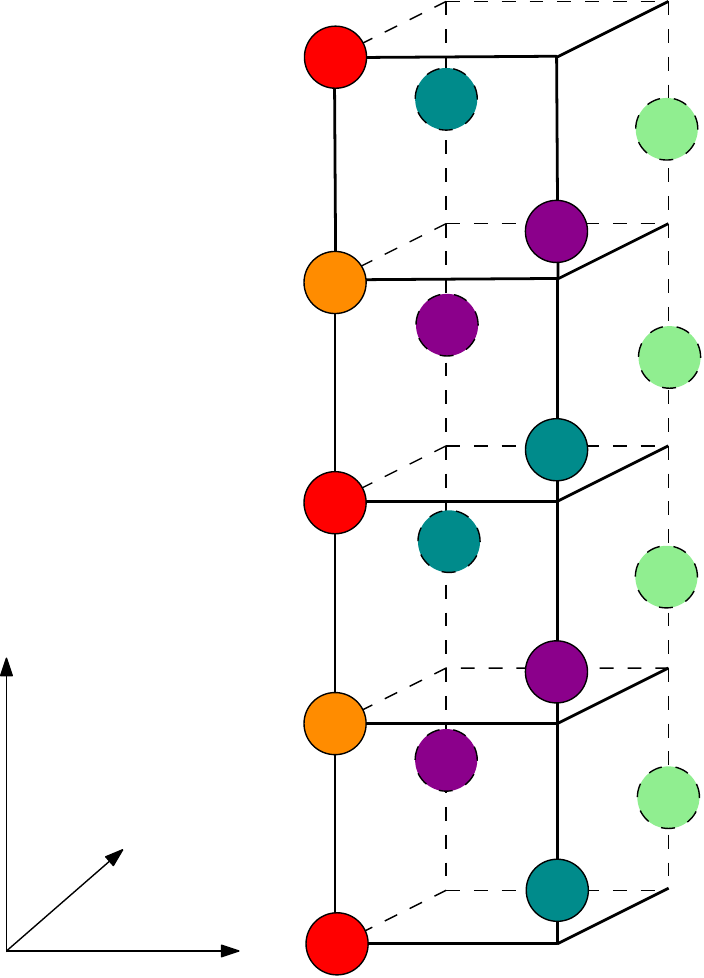}
 \put(23,5){\tiny $\mathbf a$}
 \put(-0.5,34.5){\tiny $\mathbf c$}
 \put(13,12){\tiny $\mathbf b$}
 \put(37.6,-0.60){\tiny 1}
 \put(60,10){\tiny 2}
 \put(70,21.2){\tiny 3} 
 \put(49,21){\tiny 4}
 \put(28,28){\tiny 5}
 \end{overpic}}
 \hfill
 \subfloat[]{\label{fig:mn4si7-atom1-bands}\includegraphics[width=0.38\textwidth]{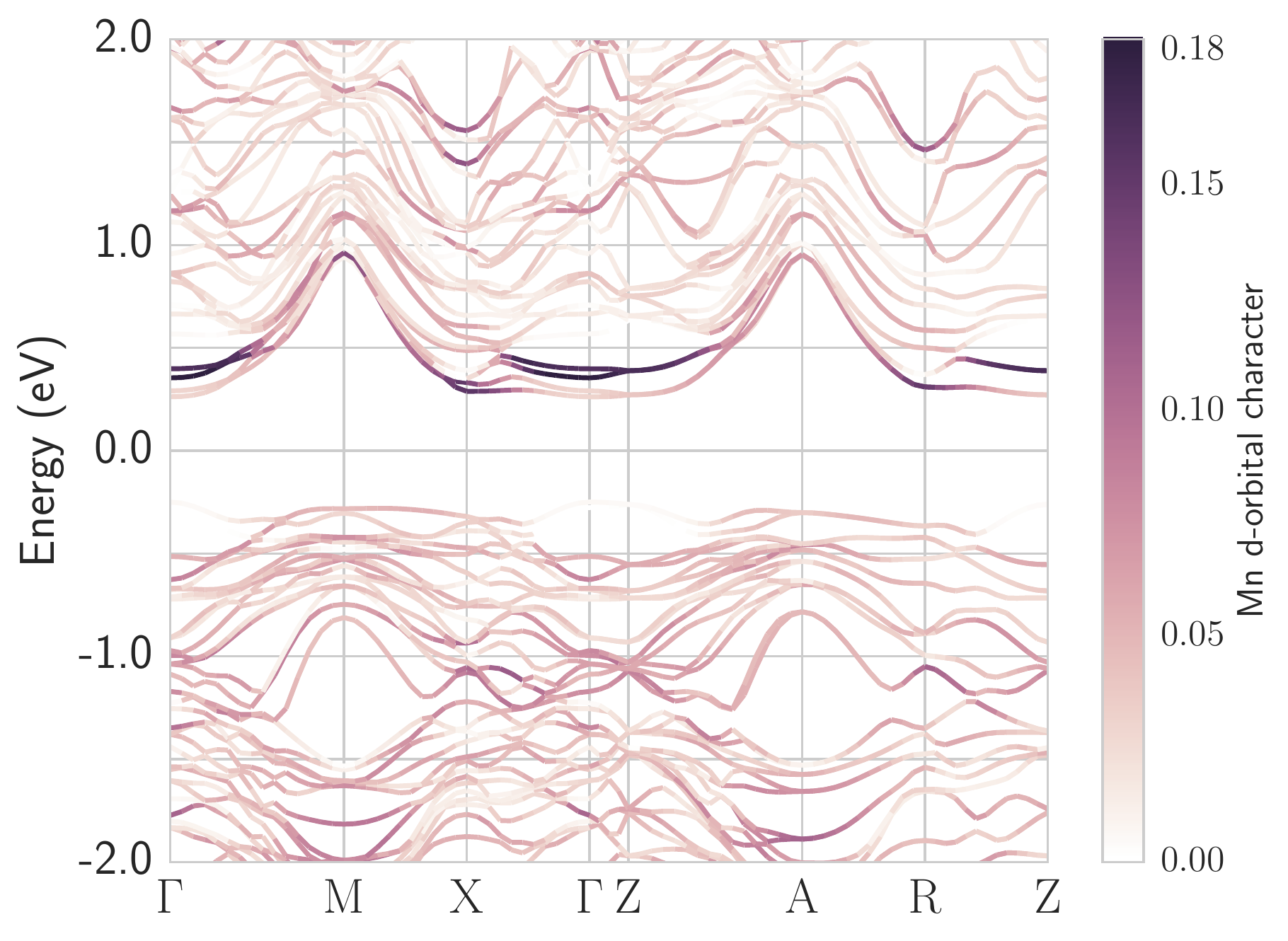}} 
 \hfill
 \subfloat[]{\label{fig:mn4si7-atom5-bands}\includegraphics[width=0.38\textwidth]{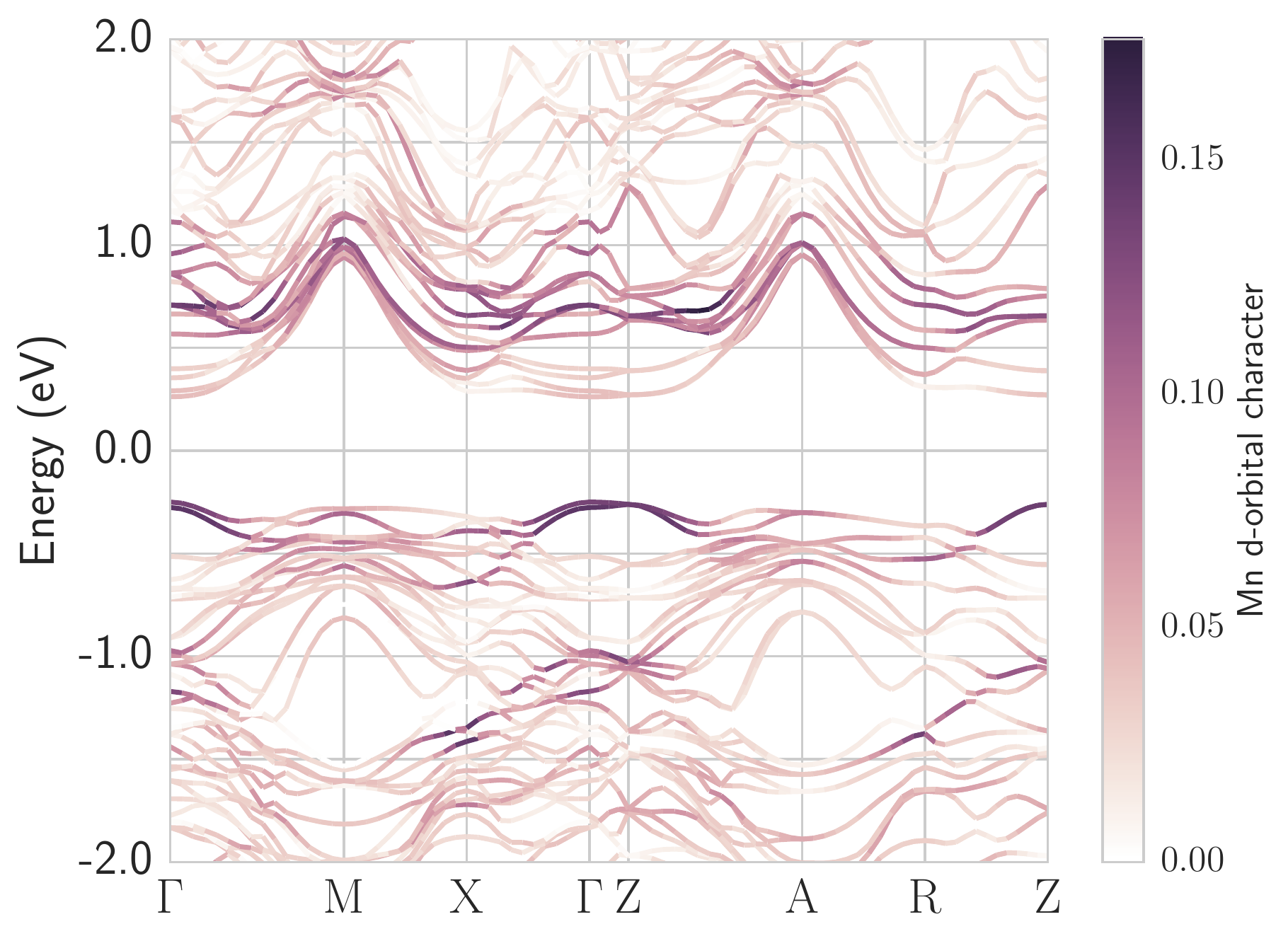}} 
   \caption{(a) The primitive cell for Mn$_4$Si$_7$ highlighting the different kinds of Mn sites using colour. The different Mn atoms in the first quarter of the cell are labelled 1-5. Circles with dashed outlines are inside the page and the cuboid extends to half a lattice vector along both the $a$ and $b$ directions. Figures (b) and (c) denote the contribution of the d-orbitals from Mn atom at (0.0, 0.0, 0.0) and the one at (0.0, 0.0, 0.25) to the band structure, respectively. The The hole like bands at the $\Gamma$ point mostly come from atom 5. The color bar in (b) and (c) indicates the weight of the particular atom and orbital type in a given band. The Fermi level is at zero.\label{fig:mn4si7-atom-contributions}}
\end{figure*}

To further understand the band structure of \ce{Mn4Si7}, we looked at the contribution of individual \ce{Mn} sites to the bands. The unit cell of \ce{Mn4Si7} comprises of five distinct \ce{Mn} sites as shown in \cref{fig:unitcell-tight-binding-5-sites}, where the first \ce{Mn} atom is coloured red and is located at (0.0, 0.0, 0.0) and the fifth \ce{Mn} atom in orange occupies (0.0, 0.0, 0.25). In \cref{fig:mn4si7-atom1-bands,fig:mn4si7-atom5-bands} we depict the contributions to the band structure from the first \ce{Mn} atom and the fifth \ce{Mn} atom, respectively. Intriguingly, most of the contributions to the hole like bands come from the fifth Mn atom. In our structural picture of \ce{Mn4Si7} built by stacking half cells and then relaxing the positions of sterically unfavourable \ce{Si} atoms, this fifth Mn atom is equidistant from the interface at both $c=0$ and $c=0.5$ and is surrounded by \ce{Si} atoms that undergo the least relaxation.

This discrepancy in the relative contribution of the \ce{Mn} atoms highlights the important role played by the \ce{Si} atoms in shaping the electronic band structure near the Fermi level. In the next subsection, we review some earlier results on the origin of the gap in the NCL compounds when there are 14 electrons per transition metal atom in the context of the HMS. 

\subsection{The origin of gap in NCLs - the 14 electron rule}
\label{ssub:14-electron-gap}
Another common feature of a wide variety of NCLs (T$_t$E$_m$) is that most of them exhibit a gap in the band structure when there are 14 electrons per transition metal atom\citep{Pearson1970,Lu2002}.  \citet{Fredrickson2004a} elucidated the origin of this gap with an analysis that combined DFT, H{\"u}ckel theory, and the bonding between the transition metal atom and the main group element for a prototype compound with a \ce{TiSi2} structure --\ce{RuGa2}. Much as in the case of \ce{Mn4Si7}, the DOS near the Fermi level is dominated by the transition metal atom \ce{Ru} in \ce{RuGa2}. In \ce{RuGa2}, the \ce{Ru} atom is surrounded by a honeycomb of \ce{Ga} atoms with a pair of \ce{Ga} atoms lying both above and below the honeycomb plane. This structural motif underlies the analysis of the bonding between the \ce{Ru} $d$ and \ce{Ga} $sp$ orbitals in \ce{RuGa2}. The gap at 14 electrons per T atom is related to the occupation of 14 orbitals per \ce{Ru} atom. While the nature of the 14 orbitals and their distribution among \ce{Ru} d, \ce{Ru}-\ce{Ga} bonding and \ce{Ga}-\ce{Ga} bonding vary depending on the $k$-point in question, there is always a gap at this electron number. 

The local environment of the fifth Mn atom in \ce{Mn4Si7} is very similar to that of \ce{Ru} in \ce{RuGa2}, with the \ce{Mn} atom surrounded by a honeycomb plane of \ce{Si} atoms and a pair of \ce{Si} atoms extending out of the honeycomb plane in either direction. As discussed in the previous section, the \ce{Si} atoms surrounding the fifth Mn atom are the ones undergoing the least structural distortion while \ce{Mn4Si7} is built from \ce{MnSi2} and highlight how the structural underpinning of the 14 electron rule is carried over to \ce{Mn4Si7}. 

Later work on \ce{RuGa2}\citep{Yannello2014} suggests that the 14 electron rule is an instance of a more general $18-n$ rule for intermetallics compounds between a transition metal T and a main group element E. $n$ is identified to be the number of T-T bonds per T atom that are mediated by an E atom which in the case of a \ce{TiSi2} structure as in \ce{MnSi2} and \ce{RuGa2} is four, thereby giving rise to the 14 electron rule. 

In the case of the HMS, only \ce{Mn4Si7} has the requisite 14 electrons per \ce{Mn} atom -- 7 electrons from each \ce{Mn} atom and 4 electrons from each \ce{Si} atom. In addition to explaining the band gap, the 14 electron rule also sheds light on why the other HMS are degenerate semiconductors. The electron count for each of the other HMS is a single electron short of the magic count of 14 electrons giving rise to a p-type band structure. Experimentally, the hole concentration is temperature independent at low temperature \citep{nishida1972semiconducting} and ranges\citep{nishida1972semiconducting,Teichert1996} from $7 \times 10^{20}$ to $3 \times 10^{21}$. According to the 14 electron rule, the hole concentration is simply the number of holes in the unit cell and is summarized for the HMS in \cref{table:electron-count-HMS}.

\begin{table}[h]
\begin{ruledtabular}
\begin{tabular}{lccc}
& \ce{Mn11Si19} & \ce{Mn15Si26} & \ce{Mn27Si47}\\
Holes per unit cell & 4 & 2& 4 \\
Unit cell volume ($10^{-21}$ cm$^3$) & 1.46 & 2.0 & 3.61\\
Hole density ($10^{21}$ cm$^{-3}$) & 2.74  & 1.0 & 1.11\\
\end{tabular}
\end{ruledtabular}
\caption{Using the 14 electron rule, we can estimate the carrier density (holes) for the HMS compounds. The holes per unit cell is obtained from the 14 electron rule and the number of formula units in the primitive cell. The hole concentration is then simply the number of holes divided by the unit cell volume.}
\label{table:electron-count-HMS}
\end{table}

\section{Effective Tight-binding model for HMS}

In previous sections, we highlighted how the electronic band structure of the HMS is mostly comprised of \ce{Mn} $d$-orbitals with the \ce{Si} $p$-orbitals playing a supporting role. In addition, we also pointed out how the structure of \ce{Mn4Si7} can be built up from half cells. In this section, we will combine these two insights to model the band structure of the HMS via tight-binding. 

The predominance of \ce{Mn} $d$-orbitals suggests that we might be able to model the electronic band structure (at least near the Fermi level) in terms of a model consisting solely of \ce{Mn} orbitals. This does not mean that the Si atoms are unimportant. The hopping integral between Mn atoms are dependent on the intervening Si atoms, and we will discuss them later. Typically in transition metal oxides, the crystal field splitting from the ions provides us with a degenerate set of local orbitals on the transition metal atom, but in the HMS, the mismatch between the periods of the \ce{Si} and \ce{Mn} atoms means that we do not have a regular local coordination for the \ce{Mn} atoms. We consider instead, an effective model for the band structure of the HMS with two spherically symmetric orbitals on each \ce{Mn} site. In the next subsection, we elaborate on the construction of this model for \ce{Mn4Si7}.

\subsection{Tight-binding model for \ce{Mn4Si7}}
\label{sec:Tight_binding_model_with_five_sublattices}
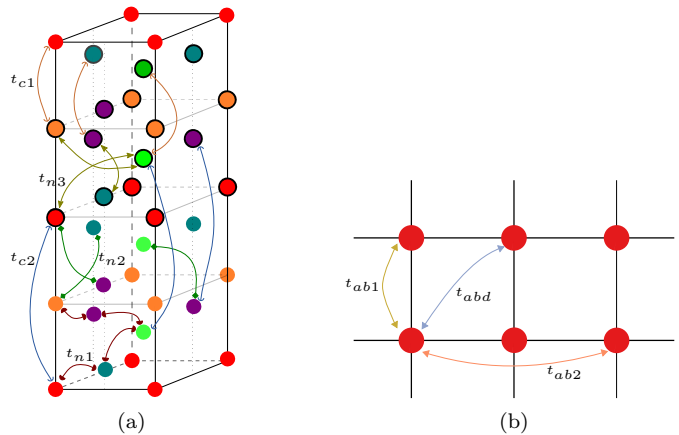
\begin{figure}[h!]
  \centering
  \subfloat[]{\label{fig:TB-out-of-plane}%
  \begin{overpic}[scale=0.45]{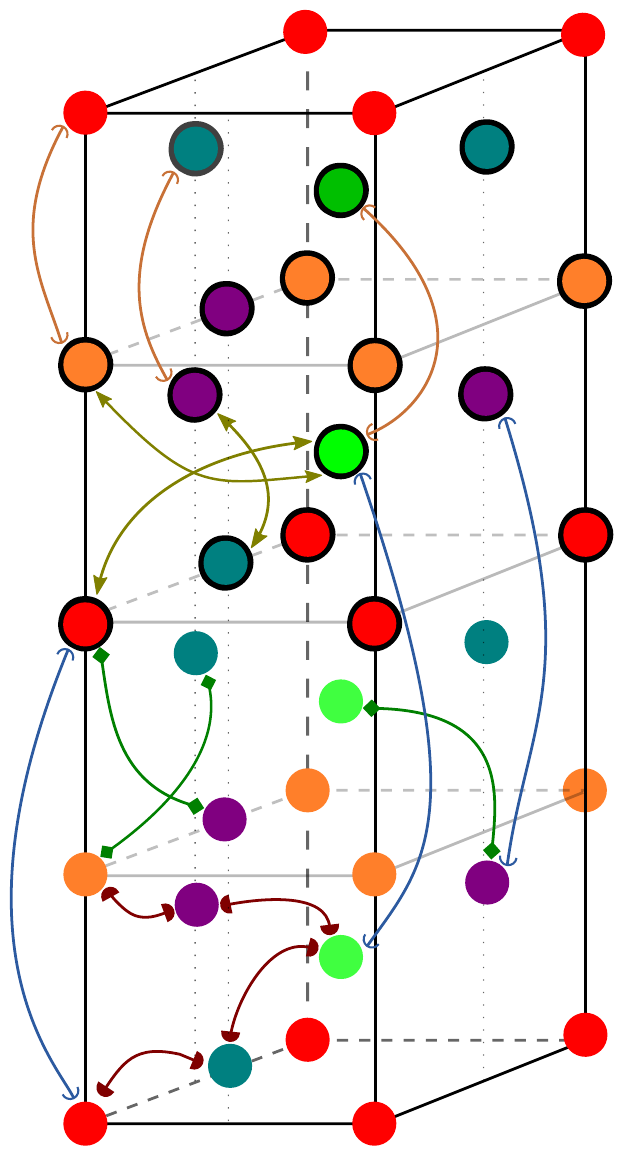}
  \put(10,11){\tiny $t_{n1}$}
  \put(18,35){\tiny $t_{n2}$}
  \put(3	,55){\tiny $t_{n3}$}
  \put(-4,80){\tiny $t_{c1}$}
  \put(-4,35){\tiny $t_{c2}$}
  \end{overpic}}
  \hfill
  \subfloat[]{\label{fig:TB-in-plane}%
  \begin{overpic}[width=0.5\columnwidth]{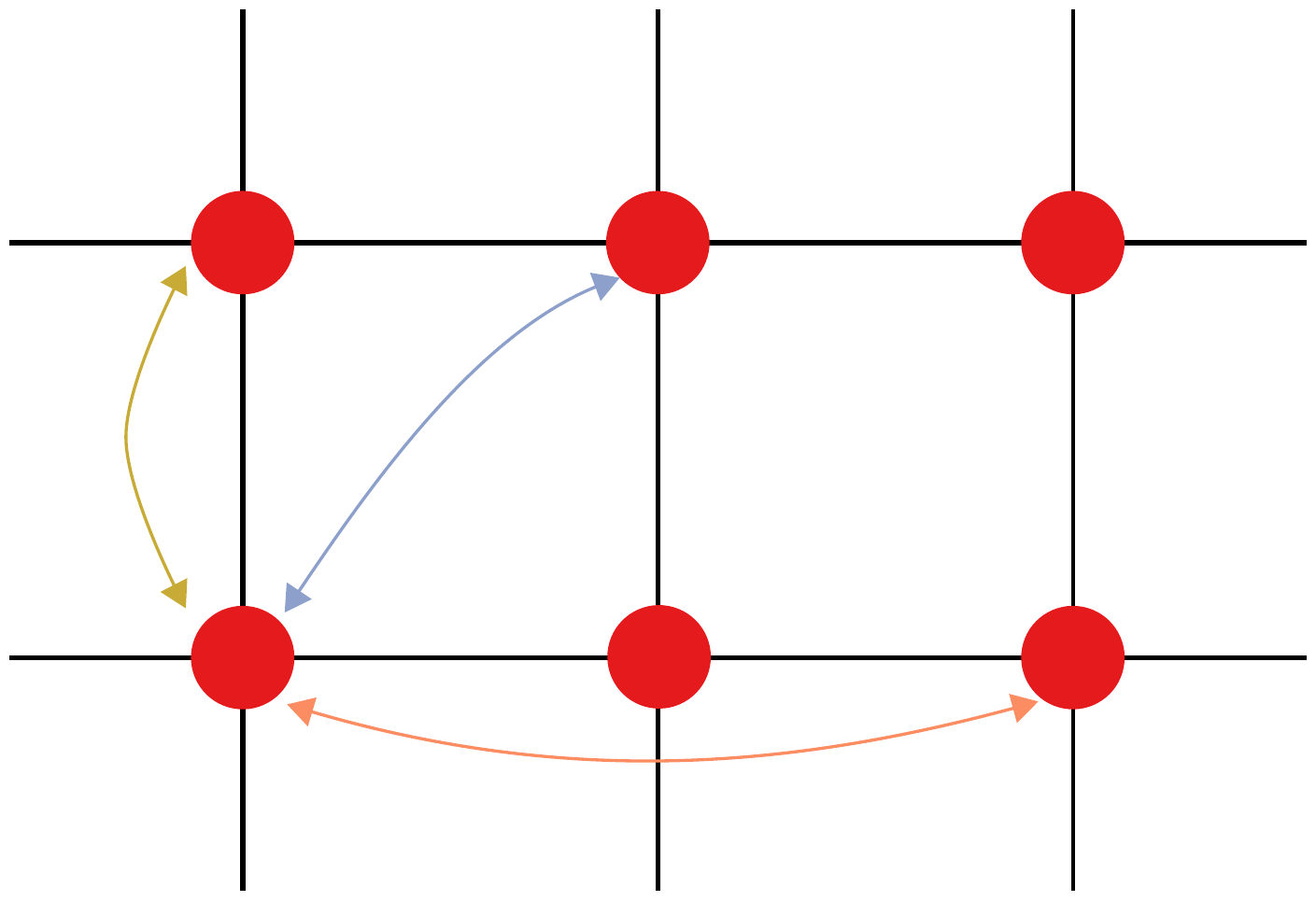}
  \put(-2,35){\tiny $t_{ab1}$}
  \put(60,8){\tiny $t_{ab2}$}
  \put(32,32){\tiny $t_{abd}$}
  \end{overpic}}
  \caption{(Color online) Tight-binding overlaps included in the minimal model. The unit cell considered for the tight-binding model is the same as in \cref{fig:unitcell-tight-binding-5-sites}, but is now split into two subunits along the $c$ axis as shown in (a). The new atoms in the unit cell that comprise the second subunit are denoted by dark enclosing circles. In (a), the overlaps $t_{n1}, t_{n2}, t_{n3}, t_{c1}$, and $t_{c2})$ are depicted with arrows. For clarity, only one kind of overlap is depicted in each quarter cell along the $c$-axis. In (b), the overlaps in the $a-b$ plane $t_{ab1}, t_{ab2}, t_{abd}$ are shown.\label{fig:TB-overlaps}}
\end{figure}

As pointed out in \cref{sub:mn4si7-structure}, the unit cell of \ce{Mn4Si7} is made up of two half-cells with five distinct \ce{Mn} atoms each. The symmetry related atoms within the unit cell are coloured the same in \cref{fig:unitcell-tight-binding-5-sites}, and \cref{fig:TB-out-of-plane} shows the unit cell once it has been split into two half-cells; the atoms belonging to the bottom half-cell denoted with colored circles and those in top half-cell denoted with dark borders for the colored circles. The actual positions of the distinct atoms in the unit cell are listed in \cref{app:mn-positions-tb}, but to simplify the our model, the fractional positions of the atoms along the $c$ axis are modified to be at integer multiples of $\tfrac{1}{16}$.

In the following sections, we will discuss a model where the structure is assumed to be of the type as shown in \cref{fig:TB-out-of-plane} with two subunits of five distinct sites each (shown in \cref{fig:unitcell-tight-binding-5-sites} in the unit cell).  

\subsubsection{Overlaps within the half-cell} 
\label{sub:overlaps_for_five_sublattice_model}

Our model consists of two orbitals on each Mn site that are separated by a gap to mimic the behaviour of the valence and conduction bands. The two orbitals have the same kind of overlaps, but we allow for the magnitudes of the overlaps to be different. The various overlaps are briefly described here, more details are in \cref{app:tb-overlaps}.

Within each half-cell, our model includes seven different types of \ce{Mn}-\ce{Mn} overlaps -- the first three are the nearest (N), the next-nearest (NN), and the third-nearest neighbour (NNN) overlaps. The next three are further neighbour overlaps for atoms located in the same $a-b$ plane  The final overlap is between Mn atoms that are directly above one another along the $c$-direction. We denote the magnitudes of these overlaps by parameters $(t_{n1}, t_{n2}, t_{n3}, t_{ab1}, t_{ab2}, t_{abd}, t_{c1})$.

\subsubsection{Overlaps between half cells} 
\label{ssub:overlaps_between_halfcells}

\begin{figure*}
  \centering
  \subfloat[]{\label{fig:tb-dos}\includegraphics[width=0.20\textwidth]{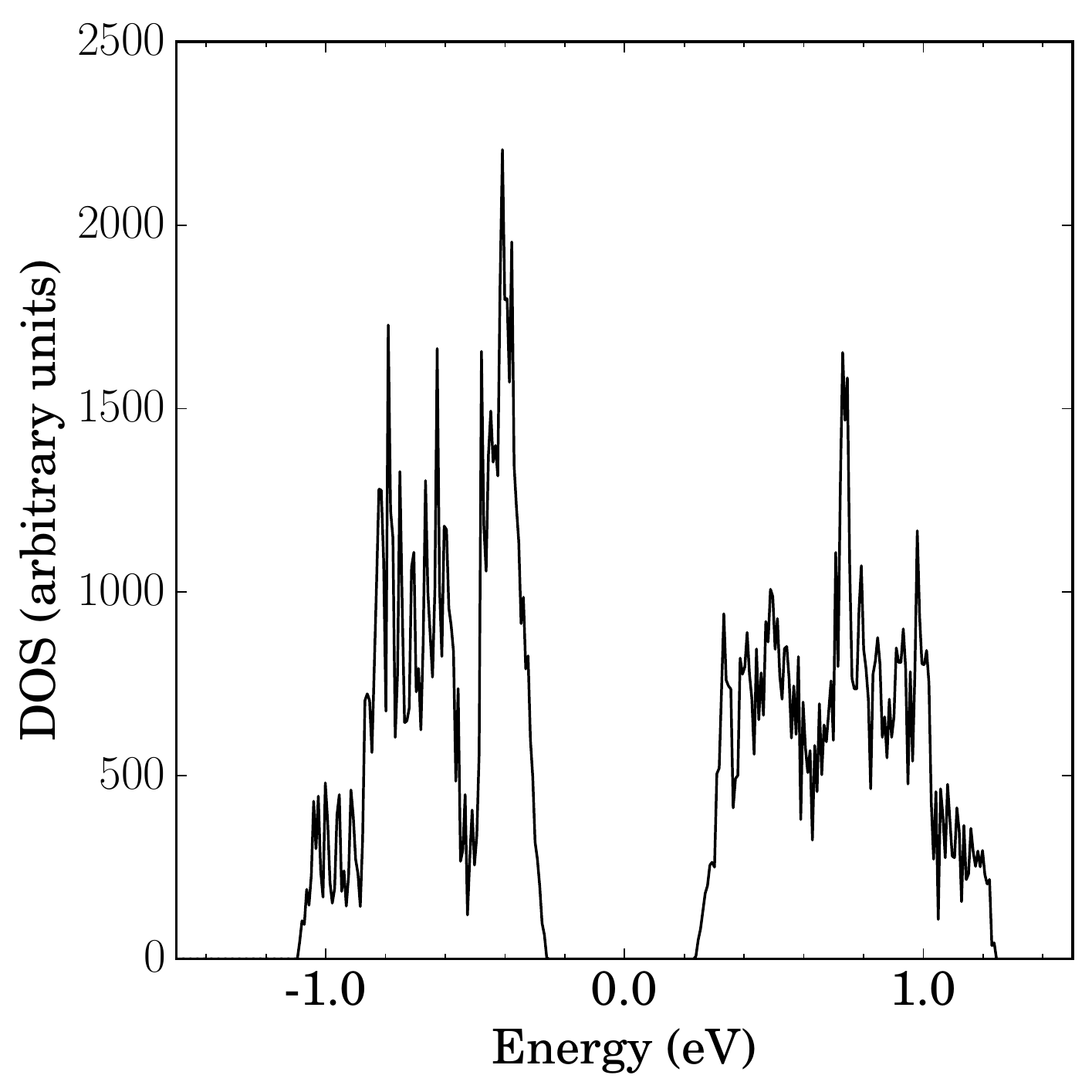}}
  \hfill
  \subfloat[]{\label{fig:mn4si7-model}\includegraphics[width=0.33\textwidth]{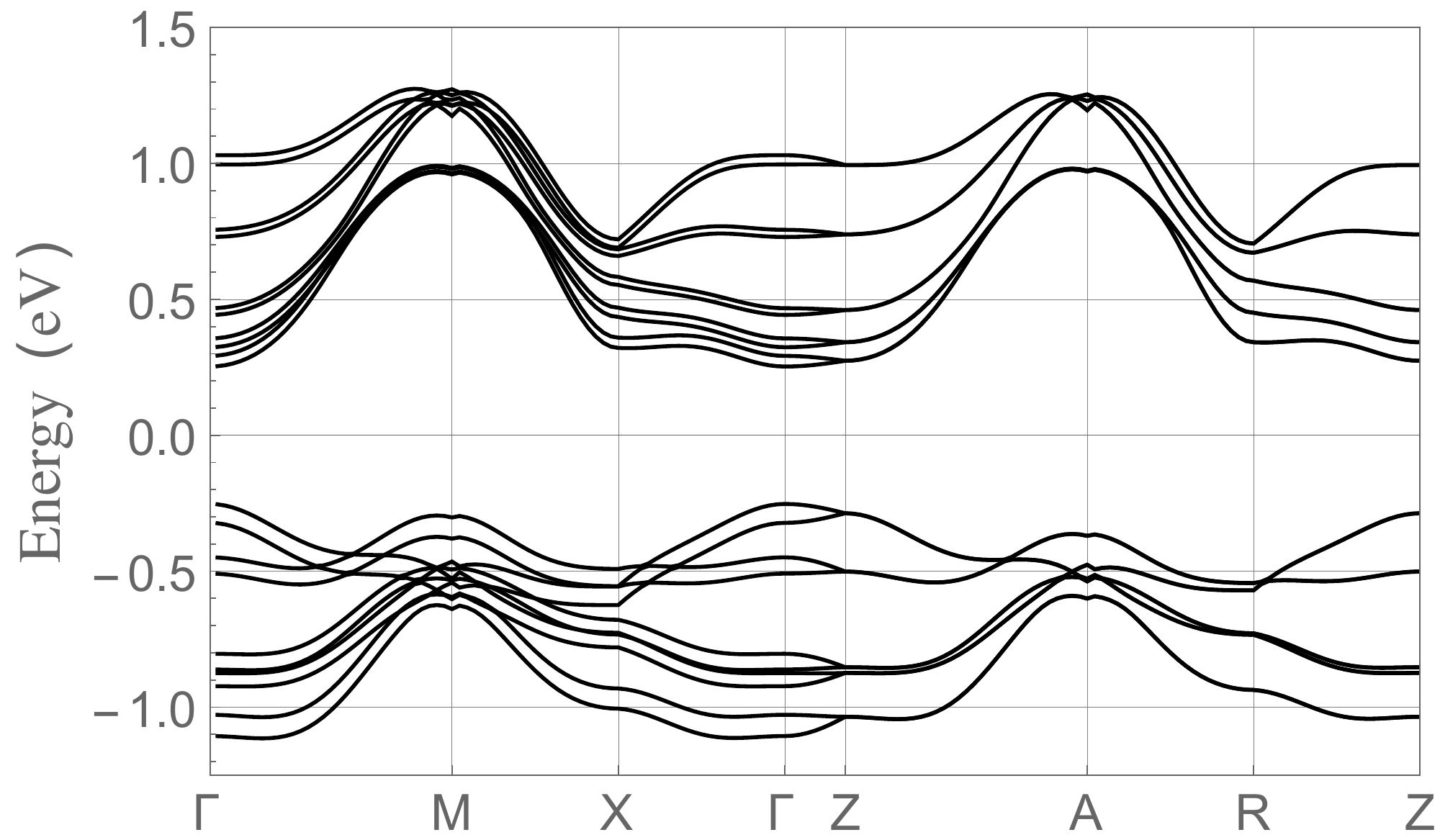}}
  \hfill
  \subfloat[]{\label{fig:mn4si7-model-no-ab}\includegraphics[width=0.33\textwidth]{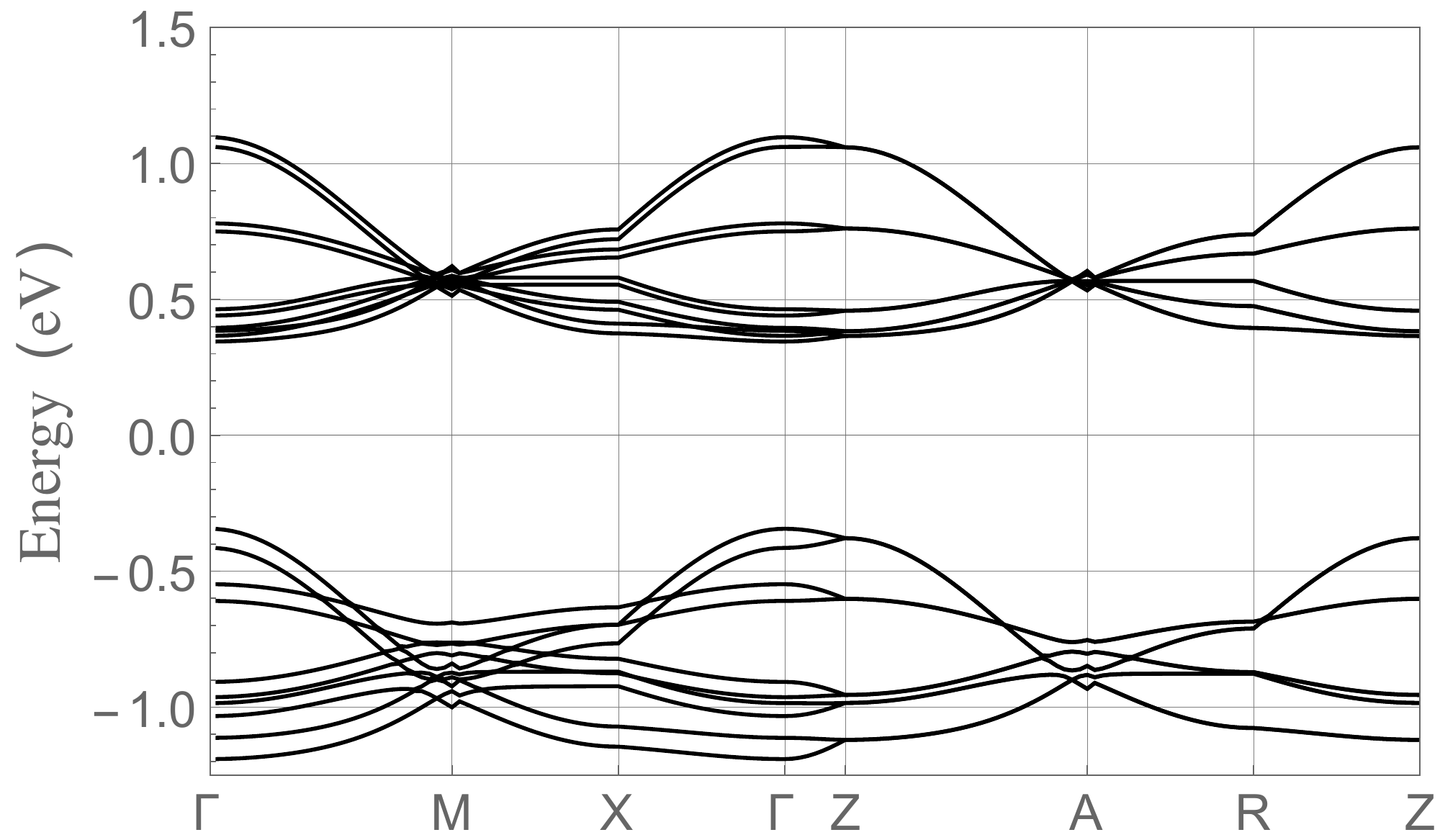}}
  \caption{Band structure and DOS for the tight-binding model. The density of states for the tight-binding model is shown in (a). For the valence bands, the tight-binding model reproduces the peak just below the Fermi level observed in the \emph{ab-initio} DOS in \cref{fig:Mn4Si7-elk-PDOS}. The two-orbital tight-binding band structures for \ce{Mn4Si7} are displayed in (b) and (c). (b) shows the band structure of the model described in \cref{sec:Tight_binding_model_with_five_sublattices} where as in (c), the further neighbour hoppings along the $a-b$ plane modified to zero. The model in (b) reproduces the DFT band structure for \ce{Mn4Si7} quite well, especially the two fold hole pockets near the $\Gamma$, and $Z$ points. On removing the further neighbour hoppings in (c), the valence bands become more dispersive near the $\Gamma$ point where as the conduction band becomes flatter. This implies that the Si atoms are important in generating the further neighbour hoppings which is required to match the DFT band structure.   \label{fig:mn4si7-model-bands}}
\end{figure*}

The two half-cell nature of the structure is incorporated into our model by supplementing the single half-cell model in the previous section with an additional hopping parametrized by $t_{c2}$ between atoms that are situated half a lattice vector along the $c$-axis. The hopping hamiltonian for each orbital is then represented in block form as
\begin{align}
     H & = \left[\begin{array}{cc} 
     h_{11} & h_{12} \\
     h_{12} & h_{11}
     \end{array}\right]
\label{eq:two-subunit-model}
\end{align}
where $H_{1}$ is the hopping for the single half-cell, and $h_{12}$ is the hopping between the subunits.

 For the valence bands, we find that a parameter set 
 \begin{align}
 (t_{n1}, t_{n2}, t_{n3}) &= (50, 5, 1.25)\\
 (t_{ab1}, t_{ab2}, t_{abd}, t_{c1}) &=  (-20, 5, 6.5 ,-25)
 \label{eq:valence-band-parameters}
 \end{align}
in units of meV for the hoppings within the subunit and an additional hopping parameter $t_{12} = 10$~meV for the hopping between half cells provides us with a good match to the DFT band structure. The hopping between the subunits is chosen to match the estimated splitting between the top two valence bands. 

The conduction bands are modeled by modifying the magnitudes of all the hoppings in \cref{eq:two-subunit-model} to obtain a Hamiltonian denoted by $\tilde{H}$. The conduction band parameters are given by the set (we use $\tilde{t}$ to distinguish between valence and conduction band parameters)
\begin{align}
(\tilde{t}_{n1}, \tilde{t}_{n2}, \tilde{t}_{n3}) &= (50, 1, 12.5)\\
(\tilde{t}_{ab1}, \tilde{t}_{ab1}, \tilde{t}_{abd},  \tilde{t}_{c1}, \tilde{t}_{12}) &= (47.5, -5, 22.5,  -1, 4)
\label{eq:conduction-band-parameters}
\end{align}

Our phenomenological model for \ce{Mn4Si7} contains two effective $s$ orbitals that are separated by a gap $\Delta=1.3~eV$ to give rise to a Hamiltonian of the form 
\begin{align}
H_{TB} & = \left[\begin{array}{cc}
H & 0 \\
0 & \Delta \1 + \tilde{H}
\end{array} \right]
\label{eq:mn4si7-model}
\end{align}

The band structures of our model is shown in \cref{fig:mn4si7-model-bands}. Considering the relative simplicity of our tight-binding model, the agreement between our band structure (which is constructed to model the bands near the Fermi level) and the one from DFT is remarkable. We are able to reproduce most of the features of the band structure for the valence bands namely the two fold hole pocket at the $\Gamma$ point and the degeneracies at the $Z$ and the $A$ points. The conduction bands near the Fermi level are also well reproduced 

The set of parameters employed to obtain this fit needs some elaboration given that we only considered overlaps between \ce{Mn} sites. The validity of the 14 electron rule in \ce{Mn4Si7} and the contributions of the fifth Mn atom to the bands near the Fermi level are both due to the bonding between Si and Mn atoms. While further calculations based on methods such as projections to Wannier orbitals are needed to underpin the exact contribution of the Si atoms to the band structure, a few key aspects can be readily inferred from the parameter set detailed above. Given that the \ce{Mn} atoms are composed of 3$d$ orbitals which are quite localized, we would expect that the direct overlaps between the \ce{Mn} $d$ orbitals would be quite small and this is indeed reflected in our tight-binding parameters where the largest overlap is only about 50~meV. The intervening Si atoms are expected to play a role in the surprisingly large further neighbour overlaps in the $a-b$ plane which are comparable to the nearest-neighbour overlaps\footnote{These hoppings are crucial for obtaining a good fit within our model. Also see Chapter 5 in \citep{VijayShankar2016perspectives}}. 

\cref{fig:mn4si7-model-bands} compares the band structure of the model for \ce{Mn4Si7} with no further neighbour hoppings ($t/\tilde{t}_{ab1, ab2, abd} = 0$). As the magnitudes of these terms are increased in the model, the effective mass of the holes in the valence band increases (valence bands become flatter) and that of the electrons in the conduction decreases near the $\Gamma$ point. In addition, the valence and conduction bands are pushed further down at the $M$(and $A$) points. Applying pressure along the $c$-axis should decrease the overlaps between orbitals in the $a-b$ plane compared to the overlaps along $c$ leading to decrease in effective mass for the valence band holes and increase for the conduction band electrons. 

\begin{figure*}
  \centering
  \subfloat[]{\label{fig:mn11si19-model}\includegraphics[width=0.4\textwidth]{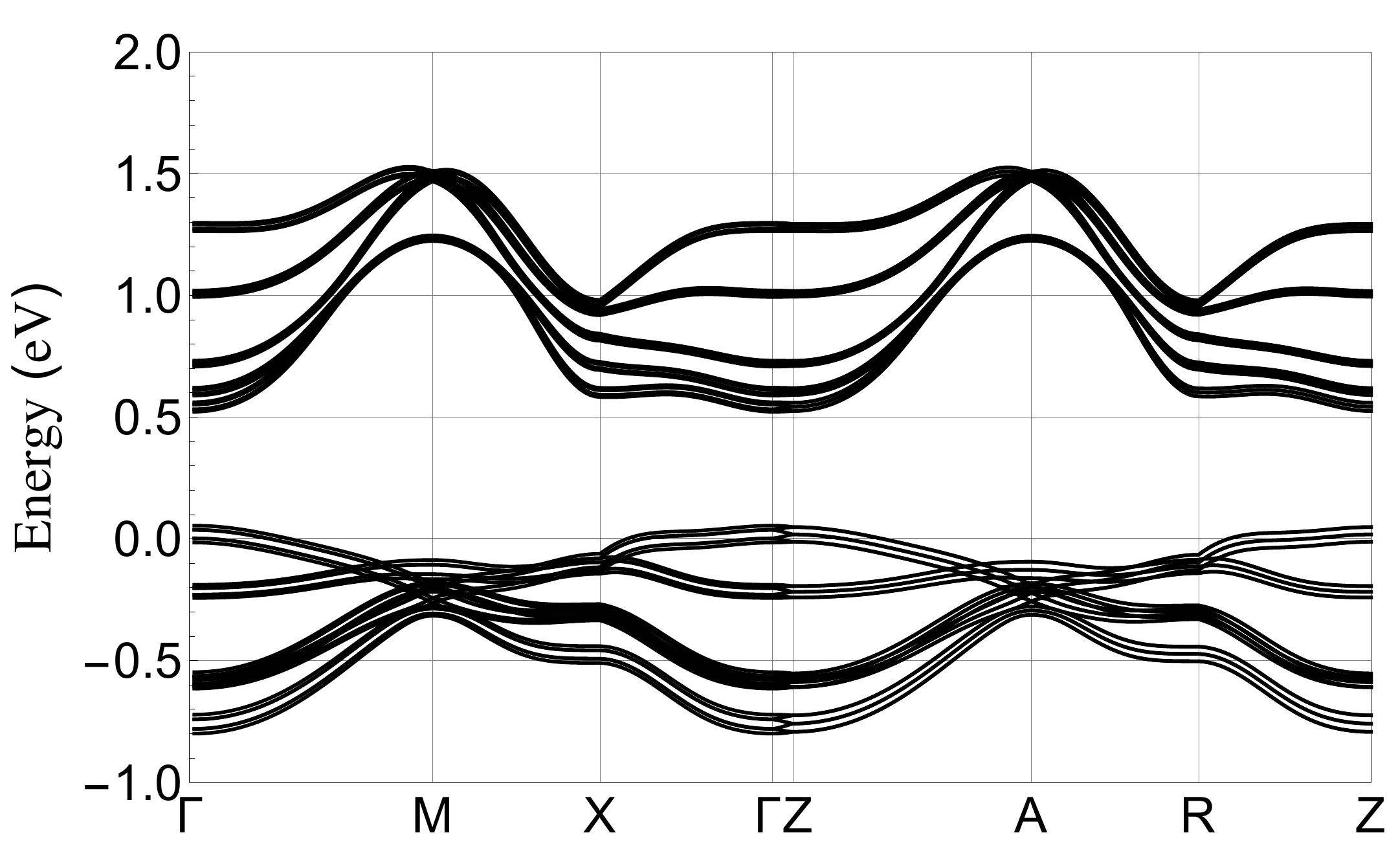}}
  \hspace{2cm}
 \subfloat[]{\label{fig:mn15si26-model}\includegraphics[width=0.4\textwidth]{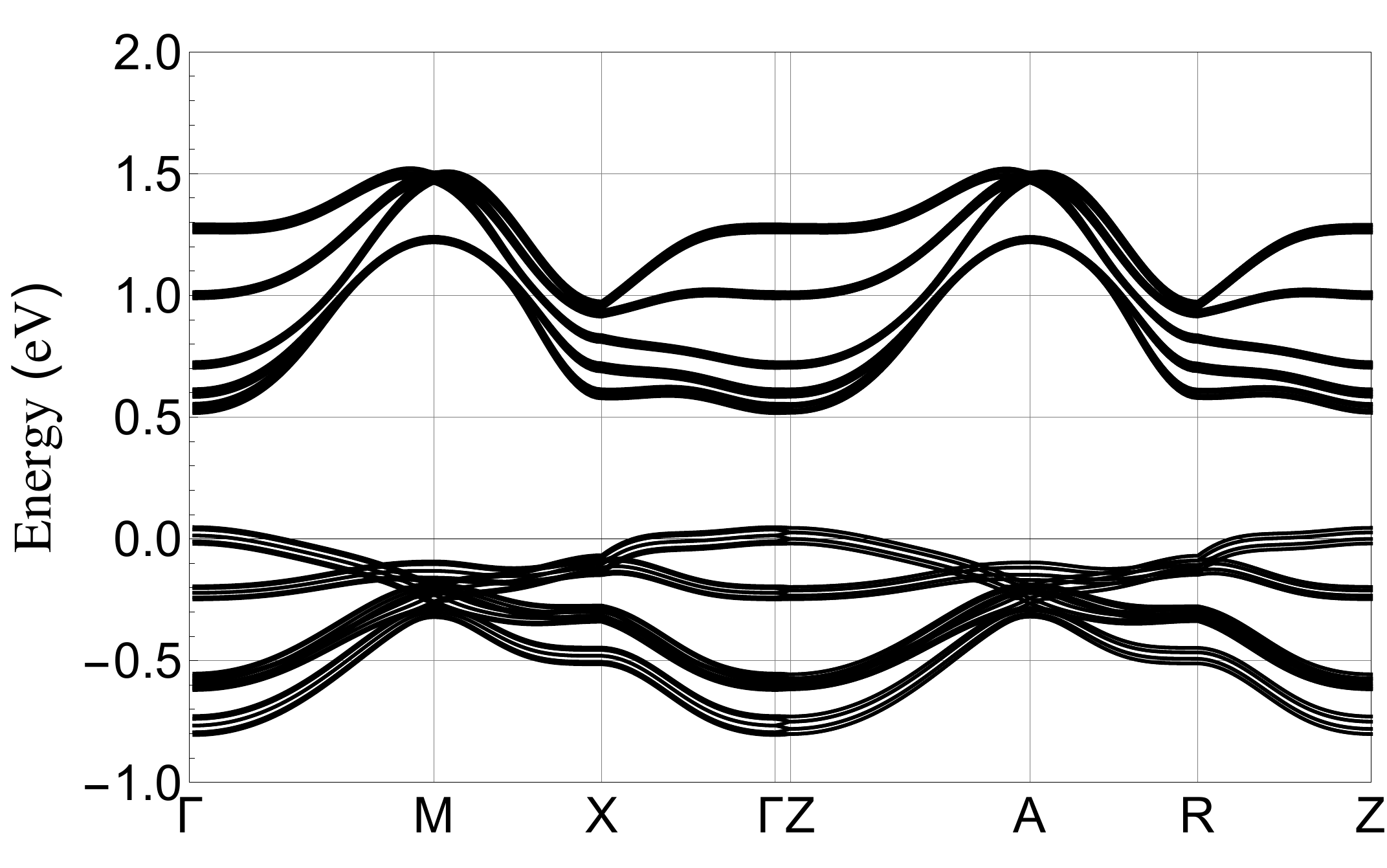}}
  \caption{Two orbital tight-binding band structures for the HMS. (a) and (b) depict the band structure obtained via stacking three and four units of \ce{Mn4Si7} to model the bands of \ce{Mn11Si19} and \ce{Mn15Si26}, respectively. The band structures for \ce{Mn11Si19} and \ce{Mn15Si26} are quite similar to the one for \ce{Mn4Si7}, except for a larger number of bands coming from the extra half cells. The 14 electron rule applied for these compounds moves the Fermi level into the valence bands. \label{fig:tb-bands}}
\end{figure*}

\subsection{Modeling the HMS band structure -- stacking a model for \texorpdfstring{\ce{Mn4Si7}}{Mn4Si7}}
\label{ssub:tb-stacking-hms}
We now proceed to discuss how we use our tight-binding model for \ce{Mn4Si7} described in the previous section, to model the other HMS. We have previously discussed how structurally these compounds can be built by stacking many halfcells of \ce{Mn4Si7}. Given that the stacking is along the $c$-axis, we employ an approach very similar to the one we described for \ce{Mn4Si7} -- add a small hopping term (equal to $t_{12}$) between the same sublattice in adjacent subunits. The stacking also increases the $c$-axis lattice constant by a factor equal to the number of \ce{Mn4Si7} units (say $n$) in the stacked unit cell and correspondingly modifies all the hopping terms described in \cref{sub:overlaps_for_five_sublattice_model} which have a $k_z$ dependence by replacing $k_z$ by $\frac{k_z}{n}$.

\ce{Mn11Si19} is modeled by stacking three \ce{Mn4Si7} unit cells or six half cells along the $c$ axis. The overlap within the half cells remain the same and those between the half cells has the same form as that discussed in \cref{sec:Tight_binding_model_with_five_sublattices}, with overlaps only between neighbouring half-cells. Since the $c$-axis lattice constant is now three times that of the orginal case for \ce{Mn4Si7}, all the $k_z$ dependent hopping terms in \cref{sub:overlaps_for_five_sublattice_model} are modified by changing $k_z \rightarrow \frac{k_z}{3}$. 

In a similar vein, \ce{Mn15Si26} is modeled by stacking four unit cells (or eight half cells) of \ce{Mn4Si7} along the $c$-axis, and modifying the $k_z$ dependence ($k_z \rightarrow \frac{k_z}{4}$) for all the hopping terms outlined in \cref{sub:overlaps_for_five_sublattice_model}.

The band structures for \ce{Mn11Si19} and \ce{Mn15Si26} are shown in \cref{fig:tb-bands} with the Fermi level chosen by ensuring that when the valence bands are filled, the 14 electron rule is obeyed. As mentioned in \cref{ssub:14-electron-gap}, both \ce{Mn11Si19} and \ce{Mn15Si26} are one electron short of having 14 electrons per \ce{Mn} atom. The Fermi level is thus within the valence bands for these two compounds. Our simple model is thus able to capture the main feature of the band structure of these compounds, namely a hole pocket near the $\Gamma$ and $Z$ points. The many half cells give rise to many bands crossing the Fermi level and the small hopping between the half cells is reflected in the small splitting between these bands. 

As pointed out in \cref{ssub:HMS-structure-stacking}, the $c$-axis lattice constant for the \ce{Mn11Si19} and \ce{Mn15Si26} is smaller than the one predicted by the $(2t -m)$ rule. For our tight-binding model, this decrease in the $c$-axis lattice constant would translate to a slight increase in the hopping along this direction, resulting in bands that would disperse slightly more along the c-axis. The overall features of the band structure will still remain the same because the $c$-axis dispersion is quite small to begin with.

\section{Discussion and Conclusions}
\label{sec:Conclusions}
Our minimal tight-binding model for the band structure of the HMS posits that the high effective mass is due to further neighbour hoppings between the \ce{Mn} atoms. This could be tested by measuring the effective mass or the Seebeck coefficient under pressure applied along the $c$-axis. Under this kind of pressure,  the $a-b$ plane overlaps are expected to decrease relative to the other overlaps and should present itself as a decreased effective mass for the valence bands. Tensile strain along the $c$-axis, on the other hand, increases the $a-b$ plane overlaps relative to the others and will result in a higher effective mass. The Seebeck coefficient is proportional to the effective mass through the Pisarenko relation, and straining the HMS might be a route to increasing the Seebeck coefficient and possibly also the thermoelectric figure of merit ZT. 

Compressive and tensile strain along the $c$-axis should also impact the electrical conductivity as it changes both the effective mass and the velocity of the electrons near the Fermi level. From our tight-binding band structure, tensile strain along the $c$-axis increases the $a-b$ plane hopping relative to the others resuling in an increase in both the effective mass and the electron velocity at the $\Gamma$ point, which presumably results in a higher electrical conductivity within a Boltzmann type approach. Indeed, a recent preprint\citep{Cuong2016density} that appeared while this manuscript was being prepared for publication, used density functional theory to study the effect of strain on \ce{Mn4Si7}reports that the $a-b$ plane conductivity decreases under tensile strain along the $a-b$ plane and increases under compressive strain. Although they do not consider the effect of strain along the $c$-axis on the $a-b$ plane, their results are broadly consistent with our tight-binding picture as we expect the $a-b$ plane hopping parameters to increase under $a-b$ plane compressive strain resulting in an increased electrical conductivity conductivity. 

Another report describing the preparation of \ce{MnSi2} phase in thin films also appeared recently\citep{Shin2015}. The measured lattice constants in \citet{Shin2015} are quite similar to that of \ce{Mn4Si7} and the measured electronic properties (Seebeck, resistivity) are similar to that of \ce{Mn11Si19}. The result is tantalizing given our model for the HMS as being built from \ce{MnSi2} cells, and it would be interesting to calculate the electronic band structure of the \ce{MnSi2} structure and compare it with that of the HMS.

In summary, using a combination of structural arguments, and electronic structure calculations, we have constructed a minimal tight-binding model for the HMS. Our main results are as follows: all the HMS can be constructed from \ce{Mn4Si7} which itself is built by stacking two half-cells of a compound with stoichiometric formula \ce{MnSi2} along the $c$ axis. The band structure of \ce{Mn4Si7} in electronic structure calculations is insulating with a gap of 0.51~eV broadly in agreement with experiments and previous theoretical calculations. A closer look at the contribution of the different \ce{Mn} atoms to the valence bands reveals that the \ce{Mn} atom that is least affected by rearrangement at the interfaces, is the majority contributor. The 14 electron rule provides an estimate for the charge density in the HMS and is remarkably close to measured values. Our minimal tight-binding model consisting of effective $s$-orbitals on \ce{Mn} atoms is able to capture most of the features of the band structure of \ce{Mn4Si7} and can be easily extended to model the band structure of the other HMS. The modeling of the thermoelectric properties of the HMS using the tight-binding model presented in this paper will be addressed in forthcoming work.

\textit{Acknowledgements -}
This work was supported by NSERC of Canada through the HEATER program and the center for Quantum Materials at the University of Toronto. Computations were performed on the GPC supercomputer at the SciNet HPC Consortium. SciNet is funded by: he Canada Foundation for Innovation under the auspices of Compute Canada; the Government of Ontario; Ontario Research Fund - Research Excellence; and the University of Toronto. VSV thanks the NSERC CREATE HEATER program for a fellowship. 

\appendix
\section{Crystal structure of \ce{Mn4Si7}}

\label{app:crystal-structure}
\begin{table}[h!]
\begin{tabular}{cccccc} 
 \multicolumn{6}{c}{Spacegroup: P$\bar{4}$c2 (no. 116)} \\ 
 \multicolumn{6}{c}{Lattice constants: a=b=5.52590~\AA{}, c= 17.5156~\AA{}}\\
 \multicolumn{6}{c}{$\alpha=\beta = \gamma = 90 ^{\circ}$}\\ \hline
  Atom & x & y & z & Wyckoff & Symmetry\\ \hline
    Mn1  &       0.00000&    0.00000&    0.00000&        2c&    -4.. \\
    Mn2 &        0.50000&    0.00000&    0.06508&        4i&     2.. \\
    Mn3 &        0.50000&    0.50000&    0.12939&        4h&     2.. \\
    Mn4 &        0.00000&    0.50000&    0.19137&        4i&     2.. \\
    Mn5 &        0.00000&    0.00000&    0.25000&        2a&   2.2 2 \\
     Si1&         0.15715&    0.20150&    0.11253&      8j&       1\\
     Si2&         0.32270&    0.84419&    0.18189&      8j&       1\\
     Si3&         0.33130&    0.33130&    0.25000&      4e&     ..2\\
     Si4&         0.34518&    0.22740&   -0.03800&      8j&       1
  \end{tabular}
 \caption{Lattice parameters, atomic coordinates, Wyckoff positions, and site symmetry for \ce{Mn4Si7} obtained from \citet{Gottlieb2003}. \label{table:mn4si7-spacegroup-positions}}
 \end{table}

\section{Computational Details}
\label{app:computational}
 The Elk package \citep{elk}, which implements the Full-Potential Linearly Augmented Plane Wave (FP-LAPW) method was used for the electronic structure calculations. The Perdew-Zunger parametrization of the local density approximation (LDA) was chosen for the exchange-correlation functional\citep{perdew1981self}.
The muffin-tin radii were automatically determined by the elk code to be 1.988 and 1.822 a.u.for Manganese and Silicon, respectively. For Brillouin zone intergration we used the \emph{highq} option in Elk, which resulted in a $k$-point grid of 5$\times$5$\times$2 corresponding to 12 points in the irreducible part of the Brillouin zone. Spin-orbit coupling and spin polarization were both ignored in our calculations. 

\section{Details of the tight-binding model}
\label{app:TB}
In this section, we provide more details of our tight-binding model including descriptions of the various terms in the tight-binding Hamiltonian. First, we describe the modified positions of the five different Mn atoms in our tight-binding unit cell and then we describe the various overlaps and the associated hopping matrices in our tight-binding model.

\subsection{Modified Mn positions}
\label{app:mn-positions-tb}
The five different Mn atoms occupy positions that are very close to multiples of $\frac{1}{16}$ as listed in \cref{table:mn4si7-spacegroup-positions}. For our tight-binding model, we shifted the atom positions to integer multiples of $\frac{1}{16}$.
\begin{table}[h!]
\begin{tabular}{cc} 
  Atom Type& Modified positions in unit cell \\ \hline
    Mn1  &    (0,0,0), (0., 0, $\tfrac{1}{2}$)       \\ \vspace{0.2em}
    Mn2 &        ($\frac{1}{2}$,0,$\tfrac{1}{16}$), (0, $\tfrac{1}{2}$, $\tfrac{6}{16}$),  ($\tfrac{1}{2}$,0, $\tfrac{9}{16}$),  (0, $\tfrac{1}{2}$, $\tfrac{15}{16}$)      \\ \vspace{0.2em}
    Mn3 &        (0, $\tfrac{1}{2}$, $\tfrac{2}{16}$), (0, $\tfrac{1}{2}$, $\tfrac{6}{16}$), (0, $\tfrac{1}{2}$, $\tfrac{10}{16}$), (0, $\tfrac{1}{2}$, $\tfrac{14}{16}$) \\ \vspace{0.2em}
    Mn4 &        (0, $\tfrac{1}{2}$, $\tfrac{3}{16}$), ($\tfrac{1}{2}$, 0, $\tfrac{5}{16}$), (0, $\tfrac{1}{2}$, $\tfrac{11}{16}$), (0, $\tfrac{1}{2}$, $\tfrac{13}{16}$) \\ \vspace{0.2em}
    Mn5 &       (0, 0 , $\tfrac{1}{4}$), (0, 0 , $\tfrac{3}{4}$) 
\end{tabular}
 \caption[Modified \ce{Mn} positions in the unit cell]{The modified positions occupied by the five different \ce{Mn} atoms in the unit cell of \ce{Mn4Si7}. Due to symmetry, there are four atoms each of types 2, 3, and 4 and two atoms each of types 1 and 5.  \label{table:mn4si7-tb-model-positions}} 
 \end{table}

\subsubsection{Tight-binding overlaps} 
\label{app:tb-overlaps}
Within each half-cell, we considered seven different types of overlaps -- the first three are the nearest (NN), the next-nearest (NN) and the third-nearest neighbour (NNN) overlaps. For each site, the nearest-neighbours are situated a fractional distance ($\pm 0.5, 0, \frac{1}{16}$) and ($\pm 0.5, 0, -\frac{1}{16}$) or with the $c$ axis fractions interchanged among the $a$ and $b$ directions. The magnitude of this overlap was denoted $t_{n1}$ and is depicted in \cref{fig:TB-out-of-plane}. 
Employing a basis $\psi_{k} = [d^{1}_{k}, d^{2}_{k},d^{3}_{k},d^{4}_{k},d^{5}_{k}]^{T}$, where $d^{\dagger i}_{k}$ creates an electron in the orbital at site $i$, the NN hopping Hamiltonian is $\sum_k \psi^{\dagger}_k h_{n1} \psi_{k}$, and the matrix $h_{n1}$ given by
\begin{align}
  h_{n1} & = \left(
\begin{array}{ccccc}
 0 & a & 0 & 0 & 0 \\
 a^*& 0 & a^* & 0 & 0 \\
 0 & a & 0 & a & 0 \\
 0 & 0 & a^* & 0 & a^*  \\
 0 & 0 & 0 & a & 0 
\end{array}
\right)
\end{align}
 where $a  = \epsilon_{1}(k_x, k_z) + \epsilon_{1}^{*}(k_y, k_z)$ with $\epsilon_{1}(k_{x/y},k_z) = 2 t_{1} \exp \left(\frac{i k_z}{16}\right) \cos\left(\frac{k_{x/y}}{2}\right)$

The next-nearest neighbours for each site are a pair of atoms situated at ($\pm 0.5, 0, \pm \frac{3}{16}$) and another pair at ($0. , \pm 0.5, \pm \frac{3}{16}$) or with $a$ and $b$ positions interchanged among these pairs. The magnitude of this $t_{n2}$ hopping is also depicted in \cref{fig:TB-out-of-plane}. The matrix $h_{n2}$ obtained for this overlap is
\begin{align}
  h_{n2} & = \left(
\begin{array}{ccccc}
 0 & 0 & 0 & c & 0 \\
 0& 0 & 0 & 0 & c* \\
 0 & 0 & 0 & c & 0 \\
 c* & 0 & c* & 0 & 0  \\
 0 & c & 0 & 0 & 0 \\
\end{array}
\right)
\end{align}
where $c  = \epsilon_{2}(k_y, k_z) + \epsilon_{2}^{*}(k_x, k_z)$  with 
$\epsilon_{2}(k_{x/y},k_z) = 2 t_{2} \exp \left(\frac{i 3 k_z}{16}\right) \cos\left(\frac{k_{x/y}}{2}\right)$. 

The third-nearest neighbours involved in the overlap $t_{n3}$ for each site are at ($\pm 0.5, \pm 0.5, \pm 0.5$) along the [111] direction and give rise to a hopping matrix $h_{3}$ with the form
\begin{align}
  h_{n3} & = \left(
\begin{array}{ccccc}
 0 & 0 & d & 0 & 0 \\
 0& 0 & 0 & d & 0 \\
 d& 0 & 0 & 0 & d \\
 0 & d & 0 & 0 & 0  \\
 0 & c & d & 0 & 0 \\
\end{array}
\right)
\end{align}
 where $d  = \epsilon_{3}(k_x, k_y, k_z) = 8 t_{3} \cos\left(\frac{k_{x}}{2}\right)\cos\left(\frac{k_{y}}{2}\right)\cos \left(\frac{k_z}{8}\right)$

 \cref{fig:TB-out-of-plane} also depicts the overlap ($t_{c1}$ from an atom to another one directly above it along the $c$ axis at a distance $(0, 0, \frac{1}{4})$. The matrix $h_{c1}$ is given by
\begin{align}
  h_{c1} &= \left(\begin{array}{ccccc}
  0 & 0 & 0 & 0 & b \\
  0& 0 & 0 & 2b & 0 \\
  0 & 0 & 2b & 0 & 0 \\
  0 & 2b & 0 & 0 & 0  \\
  b & 0 & 0 &  0& 0 \\
  \end{array}\right)
  \end{align}
 where $b = \epsilon_{z1}(k_z) = 2 t_{c1}  \cos\left(\frac{k_{z}}{4}\right)$.

The next three overlaps are overlaps between atoms situated in the same $a-b$ plane and are labelled $t_{ab1}$ and $t_{ab2}$. $t_{abd}$ and were depicted in \cref{fig:TB-in-plane}. The overlap between atoms situated a lattice vector apart in the $a-b$ plane, which we call h$_{ab1}$ is
\begin{align}
    h_{ab1} & = \left(
\begin{array}{ccccc}
 e & 0 & 0 & 0 & 0 \\
 0 & 2 e & 0 & 0 & 0 \\
 0 & 0 & 2e & 0 & 0 \\
 0 & 0 & 0 & 2e & 0  \\
 0 & 0 & 0 & 0 & e \\
\end{array}
\right)
\end{align}
where $ e  = \epsilon_{ab1}(k_x, k_y) =  2 t_{ab1 } \left( \cos (k_{x}) + \cos (k_{y}) \right)$

The overlap between atoms that are at a distance ($\pm 1, \pm 1, 0$) apart was denoted $t_{abd}$, and the associated matrix $h_{abd}$ is given by
\begin{align}
h_{abd} &= \left(
\begin{array}{ccccc}
 D& 0 & 0 & 0 & 0 \\
 0 & 2 D & 0 & 0 & 0 \\
 0 & 0 & 2 D & 0 & 0 \\
 0 & 0 & 0 & 2 D & 0  \\
 0 & 0 & 0 & 0 & D \\
\end{array}
\right)
\end{align}
where $D  = \epsilon_{D}(k_x, k_y) =  4 t_{D} \left( \cos (k_{x})\cos (k_{y}) \right)$. 

The last overlap within a half-cell is between atoms situated two lattice vectors apart in the $a-b$ plane denoted $t_{ab2}$ with the matrix $h_{ab2}$ given by
\begin{align}
    h_{ab2} & = \left(
\begin{array}{ccccc}
 f & 0 & 0 & 0 & 0 \\
 0 & 2 f & 0 & 0 & 0 \\
 0 & 0 & 2 f & 0 & 0 \\
 0 & 0 & 0 & 2f & 0  \\
 0 & 0 & 0 & 0 & f \\
\end{array}
\right)
\end{align}
 where $f  = \epsilon_{ab2}(k_x, k_y) =  2 t_{ab2 } \left( \cos (2 k_{x}) + \cos (2 k_{y}) \right)$

The factor of two for some of the atoms (2, 3, and 4) is because within each subunit, there are two instances of those atoms. Including all the different kinds of overlaps, the full hamiltonian for the single subunit pictured in \Cref{fig:unitcell-tight-binding-5-sites} is 
\begin{align}
    h_{11} = (h_{n1} + h_{n2} + h_{n3} + h_{c1} + h_{ab1} + h_{abd} + h_{ab2}).
\end{align}

 The additional hopping between the same atom in different half-cells was denoted by $t_{c2}$ in \cref{fig:TB-out-of-plane} and is associated with a matrix $h_{12}$ given by
\begin{align}
h_{12} & = 2 t_{c2} \cos(\frac{k_{z}}{2})
\left(\begin{array}{ccccc}
 1 & 0 & 0 & 0 & 0 \\
 0 & 2& 0 & 0 & 0 \\
 0 & 0 & 2 & 0 & 0 \\
 0 & 0 & 0 & 2 & 0  \\
 0 & 0 & 0 & 0 & 1 
\end{array}\right) 
\label{eq:hopping-between-subunits}
 \end{align} 
\bibliography{hms-tb.bib}
\end{document}